\documentclass[structabstract]{aa}  
\usepackage{graphicx}
\usepackage{txfonts}
\usepackage{natbib}
\usepackage{multirow}

\newcommand{\teff}  {T$_\mathrm{eff}$}
\newcommand{\logg}  {$\log g$}

\newcommand{\kms}{\hbox{${\rm km}\:{\rm s}^{-1}$}}

\def\aap{A\&A}

\def\apjs{ApJS}
\def\apjl{ApJL}

\begin{document}

   \title{Abundance to age ratios in the HARPS-GTO sample with \textit{Gaia} DR2\thanks{
Based on observations collected at the La Silla Observatory, ESO
(Chile), with the HARPS spectrograph at the 3.6 m ESO telescope (ESO
runs ID 72.C---0488, 082.C---0212, and 085.C---0063).}}
\subtitle{Chemical clocks for a range of [Fe/H]}


   \author{E. Delgado Mena\inst{1}
      \and A. Moya\inst{2,3}
      \and V. Adibekyan\inst{1}
      \and M. Tsantaki\inst{1}
      \and J.~I.~Gonz\'alez Hern\'andez\inst{4,5}
      \and G. Israelian\inst{4,5} 
      \and G. R. Davies\inst{2,3}
      \and W. J. Chaplin\inst{2,3}
      \and S. G. Sousa\inst{1,6}
      \and A. C. S. Ferreira\inst{1,6}
      \and N.~C.~Santos\inst{1,6}
     }
      
\institute{
Instituto de Astrof\'isica e Ci\^encias do Espa\c{c}o, Universidade do Porto, CAUP, Rua das
Estrelas, PT4150-762 Porto, Portugal
             \email{Elisa.Delgado@astro.up.pt}
\and
School of Physics and Astronomy, University of Birmingham, Edgbaston, Birmingham, B15 2TT, UK      
\and
Stellar Astrophysics Centre, Department of Physics and Astronomy, Aarhus University, Ny Munkegade 120, DK-8000 Aarhus C, Denmark
\and 
Instituto de Astrof\'{\i}sica de Canarias,
C/ Via Lactea, s/n, 38205, La Laguna, 
Tenerife, Spain 
\and 
Departamento de Astrof\'isica, Universidad de La Laguna, 38206 La Laguna, Tenerife, Spain
\and
Departamento de F\'isica e Astronom\'ia, Faculdade de Ci\^encias, Universidade do Porto, Portugal
}     

   \date{Received ...; accepted ...}

 
  \abstract
{}
{The purpose of this work is to evaluate how several elements produced by different nucleosynthesis processes behave with stellar age and provide empirical relations to derive stellar ages from chemical abundances.}
{We derived different sets of ages using Padova and Yonsei-Yale isochrones and Hipparcos and \textit{Gaia} parallaxes for a sample of more than 1000 FGK dwarf stars for which he have high-resolution ($R \sim$\,115000) and high-quality spectra from the HARPS-GTO program. We analyzed the temporal evolution of different abundance ratios to find the best chemical clocks. We applied multivariable linear regressions to our sample of stars with a small uncertainty on age to obtain empirical relations of age as a function of stellar parameters and different chemical clocks.}
{We find that [$\alpha$/Fe] ratio (average of Mg, Si, and Ti), [O/Fe] and [Zn/Fe] are good age proxies with a lower dispersion than the age-metallicity dispersion. Several abundance ratios present a significant correlation with age for chemically separated thin disk stars (i.e., low-$\alpha$) but in the case of the chemically defined thick disk stars (i.e., high-$\alpha$) only the elements Mg, Si, Ca, and TiII show a clear correlation with age. We find that the thick disk stars are more enriched in light-\textit{s} elements than thin disk stars of similar age. The maximum enrichment of \textit{s}-process elements in the thin disk occurs in the youngest stars which in turn have solar metallicity. The slopes of the [X/Fe]-age relations are quite constant for O, Mg, Si, Ti, Zn, Sr, and Eu regardless of the metallicity. However, this is not the case for Al, Ca, Cu and most of the \textit{s}-process elements, which display very different trends depending on the metallicity. This demonstrates the limitations of using simple linear relations based on certain abundance ratios to obtain ages for stars of different metallicities. Finally, we show that by using 3D relations with a chemical clock and two stellar parameters (either \teff, [Fe/H] or stellar mass) we can explain up to 89\% of age variance in a star. A similar result is obtained when using 2D relations with a chemical clock and one stellar parameter, explaining up to a 87\% of the variance.}
{}


\maketitle
%

\section{Introduction}

The combination of precise chemical abundances with stellar ages in different stellar populations opens the door to a more insightful view on Galactic archaeology. Moreover, there is an active discussion on how the thin and thick disk components of the Galaxy should be defined, with an increasing number of works suggesting that age rather than kinematics is a better parameter to differentiate both populations \cite[e.g.,][]{haywood13,bensby14}. Therefore, a great effort to derive reliable ages in big samples of stars has been done in the previous years. The use of asteroseismic observations represents a significant advance in the derivation of accurate ages but the samples analyzed with this method are still limited in size \citep[e.g., Kepler LEGACY stars in][]{nissen17} or limited to a range of stellar parameters \citep[e.g., red giants in APOKASC and in CoRoGEE dataset,][respectively]{pinsonneault18,anders17} in order to allow for a global comprehensive analysis of the Milky Way. However, the arrival of \textit{Gaia} data will have a significant impact in the derivation of reliable ages for large sample of stars across the full Galaxy. Indeed, some large spectroscopic surveys are already taking advantage of \textit{Gaia} DR1 and DR2 to derive ages and distances, such as LAMOST \citep[e.g.,][]{tian_lamost18,yu_lamost18}, Gaia-ESO survey \citep{randich18}, GALAH \citep[e.g.,][]{buder18} and APOGEE \citep[e.g.,][]{feuillet18,fernandez_apogee18}.

Furthermore, constraining the temporal evolution of different chemical species can help to understand which nucleosynthesis channels are taking place at different ages and evaluate the relative importance of stellar yields at a given time. Moreover, the different sources producing elements (e.g., massive stars, neutron star mergers, AGB, SNeIa) are not distributed in an homogeneous way across the Galaxy. Therefore, the current abundances of a given stellar population are the outcome of several conditions that need to be accounted for to reconstruct the chemical evolution of the Galaxy. As shown in this work, the distinction of different abundance-age trends requires detailed and very precise chemical abundances that are not always possible to achieve in larger spectroscopic surveys. Therefore, although our sample is of a modest size compared to previously mentioned surveys and only covers the solar neighborhood, the high resolution and high S/N of the data will serve to make an important contribution for future models of Galactic chemical evolution (GCE). 

In this work we provide stellar ages for the HARPS-Guaranteed Time Observations (GTO) sample in order to study the temporal evolution of chemical species with different nucleosynthetic origin and to analyze the feasibility of using different abundance ratios to estimate stellar ages. The structure of the paper is the following: in Sects. \ref{sec:data} and \ref{sec:ages} we describe the sample of chemical abundances and the derivation of stellar ages. In Sects. \ref{sec:feh_age} and \ref{sec:xfe_age} we evaluate the temporal evolution of metallicity and different abundances ratios. Sect. \ref{sec:clocks} is devoted to find statistically significant correlations between age and abundance ratios with the aim of obtaining simple relations to derive stellar ages. Finally, in Sect. \ref{sec:conclusions} we present the summary of the results. 

\begin{center}
\begin{table}
\caption{Solar abundances from this work (using \logg\,=\,4.39\,dex), left column, and from DM17 (using \logg\,=\,4.43\,dex), right column.}
\label{abun_solar}
\centering
\begin{tabular}{lcc}
\hline
\noalign{\medskip} 
Element &  log (A) & log (A) \\
\noalign{\medskip} 
\hline
\hline
\noalign{\smallskip} 
\ion{Al}{I}&   6.472 & 6.470  \\
\ion{Mg}{I}&   7.584 & 7.580  \\
\ion{Si}{I}&   7.550 & 7.550  \\
\ion{Ca}{I}&   6.366 & 6.360  \\
\ion{Ti}{I}&   4.992 & 4.990  \\
\ion{Ti}{II}&   4.972 & 4.990  \\
\ion{Cu}{I}&   4.101 & 4.102  \\
\ion{Zn}{I} &  4.531 & 4.532  \\
\ion{Sr}{I} &   2.783 &  2.780 \\
\ion{Y}{II} &  2.210 & 2.224  \\
\ion{Zr}{II} &   2.647 & 2.663  \\
\ion{Ba}{II} &   2.254 & 2.259  \\
\ion{Ce}{II} &   1.603 & 1.620  \\
\ion{Nd}{II} &   1.709 & 1.726 \\
\ion{Eu}{II} &   0.654 & 0.670 \\
\noalign{\smallskip} 
\hline
\hline
\end{tabular}
\end{table}
\end{center}

\begin{center}
\begin{table*}
\caption{Stellar ages and masses obtained with \textit{Gaia} DR2 parallaxes and the PARAM interface. The V magnitude and the parallax (considering the systematics and errors as described in Sect. 3) for each star are also listed. The complete version of this table can be found in the online version.}
\label{tab:ages}
\centering
\begin{tabular}{lcccc}
\hline
\noalign{\medskip} 
Star &  $M$  & Age  & $V$ & plx \\
\noalign{\medskip} 
\hline
\hline
\noalign{\smallskip}
HD183870  &  0.780 $\pm$ 0.022  &   5.452 $\pm$  4.583 & 7.5300   &   56.546 $\pm$ 0.080    \\
HD115617  &  0.918 $\pm$ 0.034  &   7.309 $\pm$  3.775 & 4.7400   &  117.603 $\pm$ 0.340    \\
HD17439   &  1.012 $\pm$ 0.034  &   3.455 $\pm$  2.577 & 8.6300   &   17.559 $\pm$ 0.041    \\
HD78612   &  0.943 $\pm$ 0.020  &  10.214 $\pm$  0.804 & 7.1500   &   24.243 $\pm$ 0.076    \\
HD201422  &  0.954 $\pm$ 0.032  &   2.109 $\pm$  1.983 & 8.5400   &   19.791 $\pm$ 0.087    \\
HD29980   &  1.110 $\pm$ 0.030  &   1.243 $\pm$  1.115 & 8.0300   &   19.829 $\pm$ 0.068    \\
\hline
\end{tabular}
\end{table*}
\end{center}

\section{Stellar parameters and chemical abundances}\label{sec:data}

The baseline sample used in this work consist of 1111 FGK stars observed within the context of the HARPS-GTO planet search programs \citep{mayor03,locurto,santos_harps4}. The final spectra have a resolution of R $\sim$115000 and high signal-to-noise ratio (45$\%$ of the spectra have 100\,$<$\,S/N\,$<$\,300, 40$\%$ of the spectra have S/N\,$>$\,300 and the mean S/N is 380). 

Precise stellar parameters for the full sample of 1111 stars within the HARPS-GTO program were homogeneously derived in \cite{sousa08,sousa_harps4,sousa_harps2}. The parameters for cool stars were revised by \citet{tsantaki13} using a special list of iron lines which was later applied to the full sample in \citet[][hereafter DM17]{delgado17}, also correcting the spectroscopic gravities. From the 1111 stars in the original sample, the derivation of parameters converged to a solution for 1059 of them. Our stars have typical \teff\ values between 4500\,K and 6500\,K and surface gravities mostly lie in the range 4\,$<$\,$\log g$\,$<$\,5 dex meanwhile the metallicity covers the region -1.39\,$<$\,[Fe/H]\,$<$\,0.55\,dex. Chemical abundances of Cu, Zn, Sr, Y, Zr, Ba, Ce, Nd, and Eu were determined under local thermodynamic equilibrium (LTE) using the 2014 version of the code MOOG \citep{sneden} and a grid of Kurucz ATLAS9 atmospheres \citep{kurucz}. For more details about the sample and the analysis we refer the reader to DM17. In that work we also provide updated values of Mg, Al, Si, Ca, and Ti using the corrected values of \logg\ and \teff\ and the EWs measured by \citet{adibekyan12}. 

In the analysis presented in DM17 we used as solar reference abundances those derived with the spectroscopic \logg\ (4.43\,dex) instead of using the corrected value considering its \teff\ (4.39\,dex, obtained with Eq. 2 of such work) as done for the full sample. The differences are minimal and well below the errors ($<$\,0.006\,dex for neutral elements and $<$\,0.018\,dex for ionized species) so they do not change the conclusions presented in DM17. However, since our aim is to obtain empirical relations to derive stellar ages through multivariable linear regressions we must consider this small difference. The chemical clocks are typically made by the subtraction of one neutral element from one ionized element (e.g., [Y/Mg]) so the difference might be slightly bigger and could produce an small offset with the empirical relations found by other authors (P. E. Nissen, private communication). Therefore, we re-derived the [X/Fe] of the full sample by using the solar reference values obtained with the corrected \logg. In Table \ref{abun_solar} the solar abundances from DM17 are compared with the new ones using \logg\,=\,4.39\,dex.

\begin{figure}
\centering
\includegraphics[width=1\linewidth]{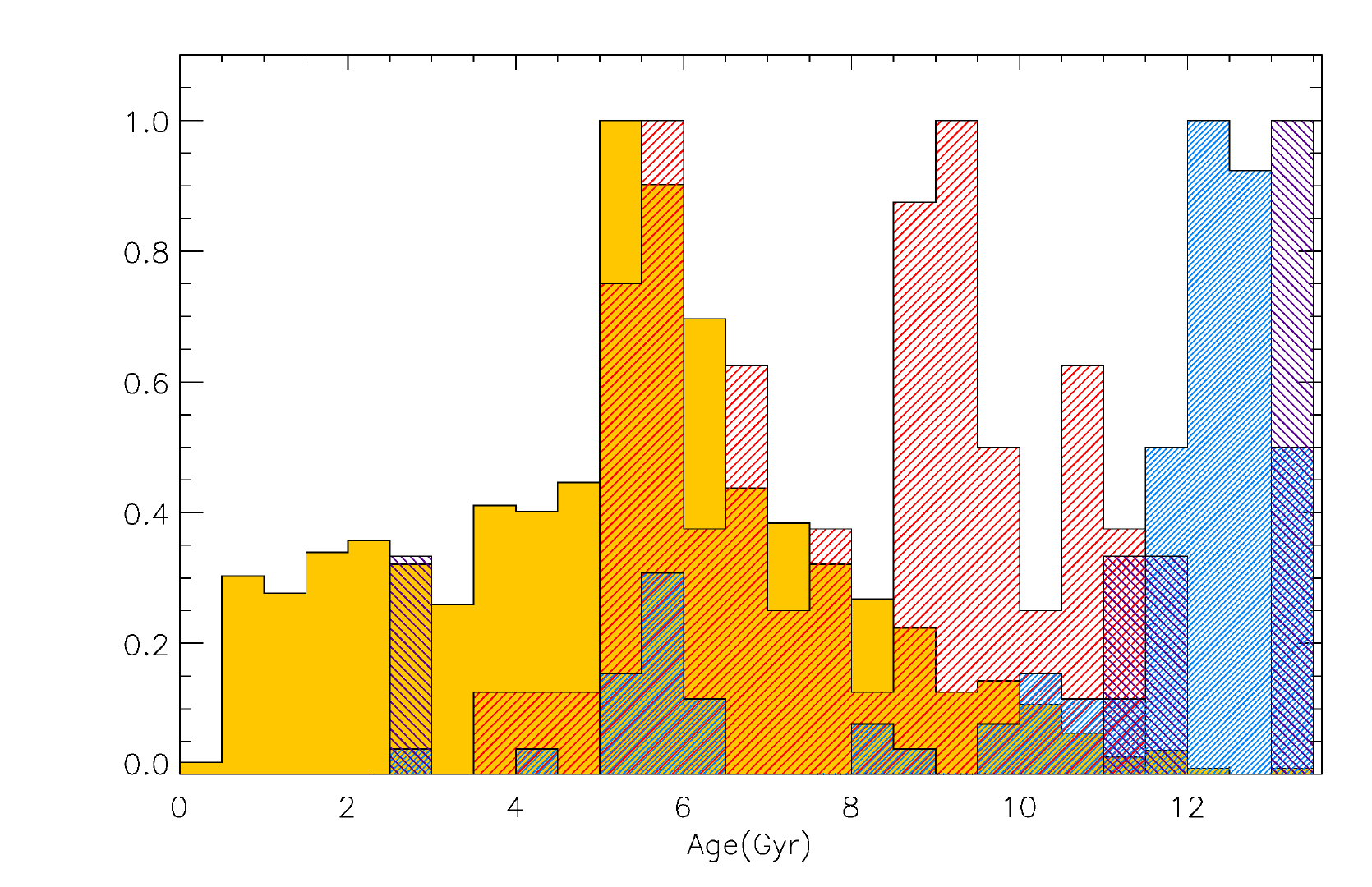}
\includegraphics[width=1\linewidth]{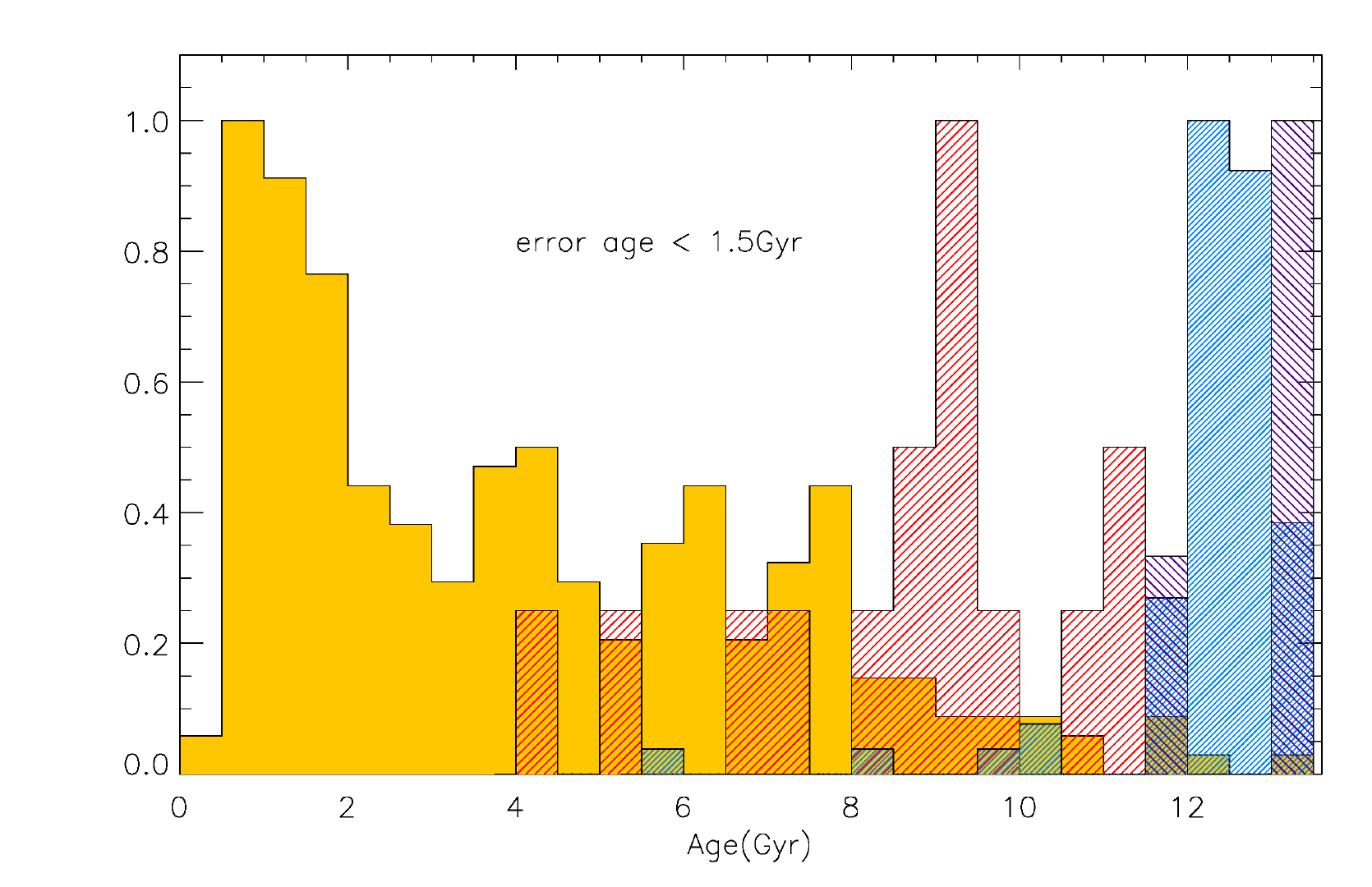}
\caption{Normalized distribution (per stellar population) of ages for the full sample (upper panel) and a subset with small errors in age (lower panel) using \textit{Gaia} DR2 parallaxes. The yellow, red, blue and purple histograms represent the thin, h$\alpha$mr, thick and halo stars, respectively.} 
\label{histo_ages}
\end{figure}

\section{Derivation of stellar ages}\label{sec:ages}

In this work we have derived three sets of masses and ages, the last two for comparison purposes: 

\begin{itemize}
 \item \textit{Gaia} DR2 parallaxes together with PARSEC isochrones
 \item Hipparcos parallaxes together with PARSEC isochrones 
 \item \teff\ and \logg\ together with Yonsei-Yale isochrones
\end{itemize}

For the first two sets we derived the masses, radii and ages with the PARAM v1.3 tool\footnote{http://stev.oapd.inaf.it/cgi-bin/param\_1.3} using the PARSEC isochrones \citep{bressan12} and a Bayesian estimation method \citep{dasilva06_param} together with the values for \teff\ and [Fe/H] from DM17, the V magnitudes from the main Hipparcos catalog \citep{perryman97} and the parallaxes from the second release (DR2) of \textit{Gaia} \citep{gaia_mission16,gaia_DR2_2018,lindegren18_gaia}, available for 1057 out of 1059 stars. All the stars in our sample have errors in parallax well below the value of the parallax itself, with most of them having an error in parallax lower than 0.1\,mas. The Bayesian inference is applied taking into account priors for the initial mass function \citep{chabrier01} and a constant Star Formation Rate. We used as a prior a maximum age of 13.5\,Gyr. No correction for interstellar reddening was considered as all stars are in close distance. We note that the errors in \teff\ and [Fe/H] that need to be input in PARAM to derive the ages should be absolute errors. Therefore, the final errors are the quadratic sum of the precision errors (those reported in DM17) and a systematic error of 60\,K and 0.04\,dex as determined by \citet{sousa_harps4}. 

The distribution of ages in the full sample can be observed in the upper panel of Fig. \ref{histo_ages} for the different populations as defined in \citet{adibekyan11,adibekyan13}. We recall that the thin and thick disk stars are defined based in the chemical separation in [$\alpha$/Fe] (being $\alpha$ the average of Si, Mg, and Ti) across different metallicities bins \citep[see Figs. 1 and 2 and Fig. 9 in][and DM17, respectively]{adibekyan11}. That work also revealed the existence of a high-$\alpha$ metal-rich (hereafter h$\alpha$mr) population, with [Fe/H]\,$>$\,-0.2\,dex and enhanced [$\alpha$/Fe] ratios with respect to the thin disk. On the other hand, halo stars are defined based on their kinematics alone.
Thick disk stars have a peak at around 12\,Gyr and high-$\alpha$ metal-rich stars are clearly older on average than thin disk stars, with two peaks around 6 and 9\,Gyr. As expected, most of the halo stars are older than 11\,Gyr. We note that if we consider those stars with small errors in age ($<$\,1.5\,Gyr) the distribution is different (see lower panel of Fig. \ref{histo_ages}), with thin disk stars peaking at younger ages because hotter stars tend to have smaller errors in parameters and thus also smaller errors in age. Also, the peak of thin disk and h$\alpha$mr stars around 6\,Gyr disappears since that was produced by cool stars with large errors in age. Finally, the majority of thick disk and halo stars with ages lower than $\sim$ 8\,Gyr are also removed due to their large errors in age.

\begin{figure}
\centering
\includegraphics[width=1\linewidth]{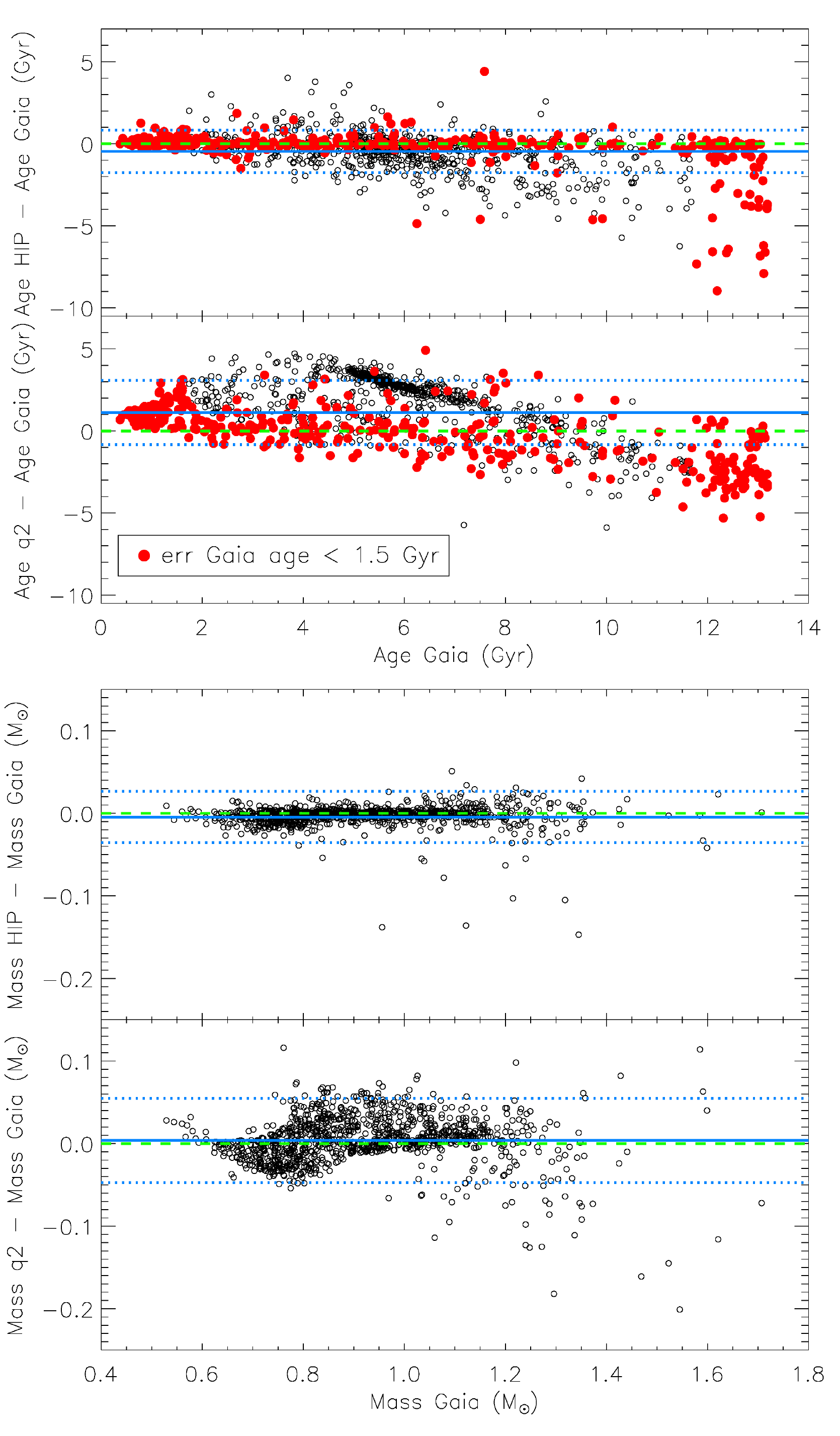}
\caption{Comparison of ages and masses obtained using \textit{Gaia} DR2 or Hipparcos parallaxes and obtained with package \textit{q$^{2}$}. The red dots are the stars with errors in age lower than 1.5\,Gyr. The green dashed lines shows the zero differences and the blue lines show the mean and standard deviation.} 
\label{comp_ages}
\end{figure}

\begin{figure}
\centering
\includegraphics[width=1\linewidth]{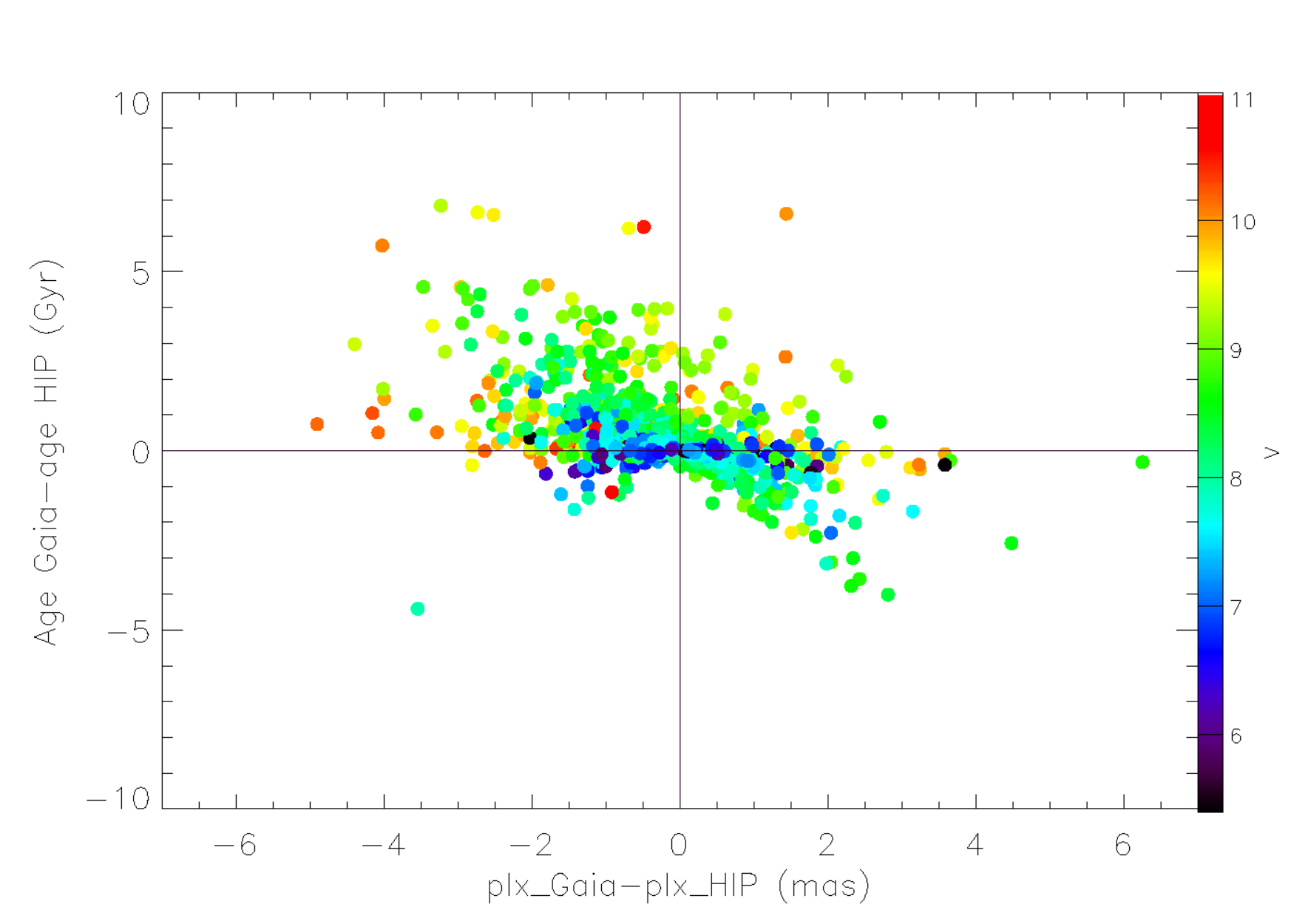}
\caption{Comparison of age differences vs parallax differences between \textit{Gaia} DR2 and Hipparcos. The V magnitudes are shown in a color scale.} 
\label{comp_ages_plx}
\end{figure}

\begin{figure}
\centering
\includegraphics[width=1\linewidth]{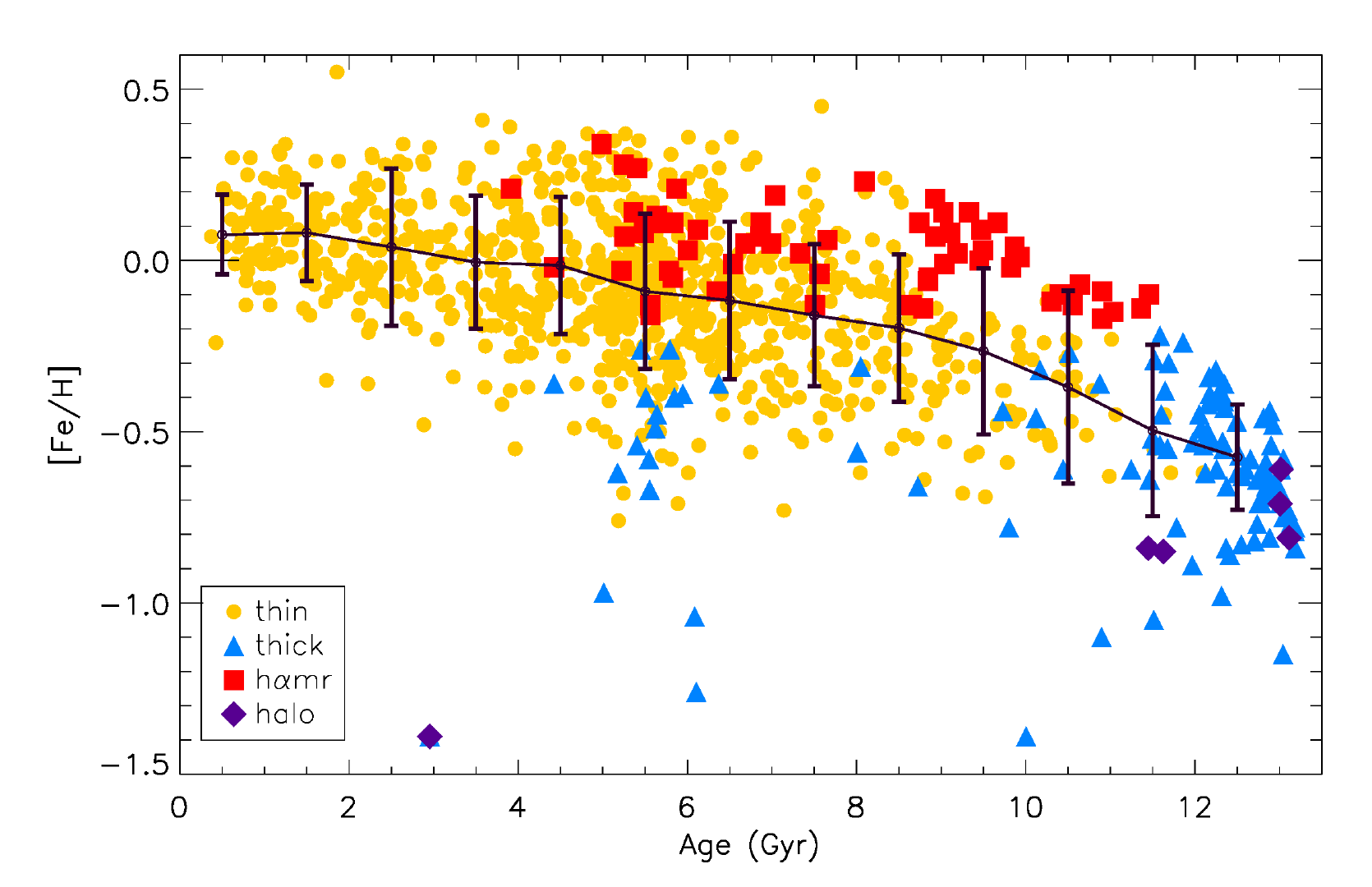}
\includegraphics[width=1\linewidth]{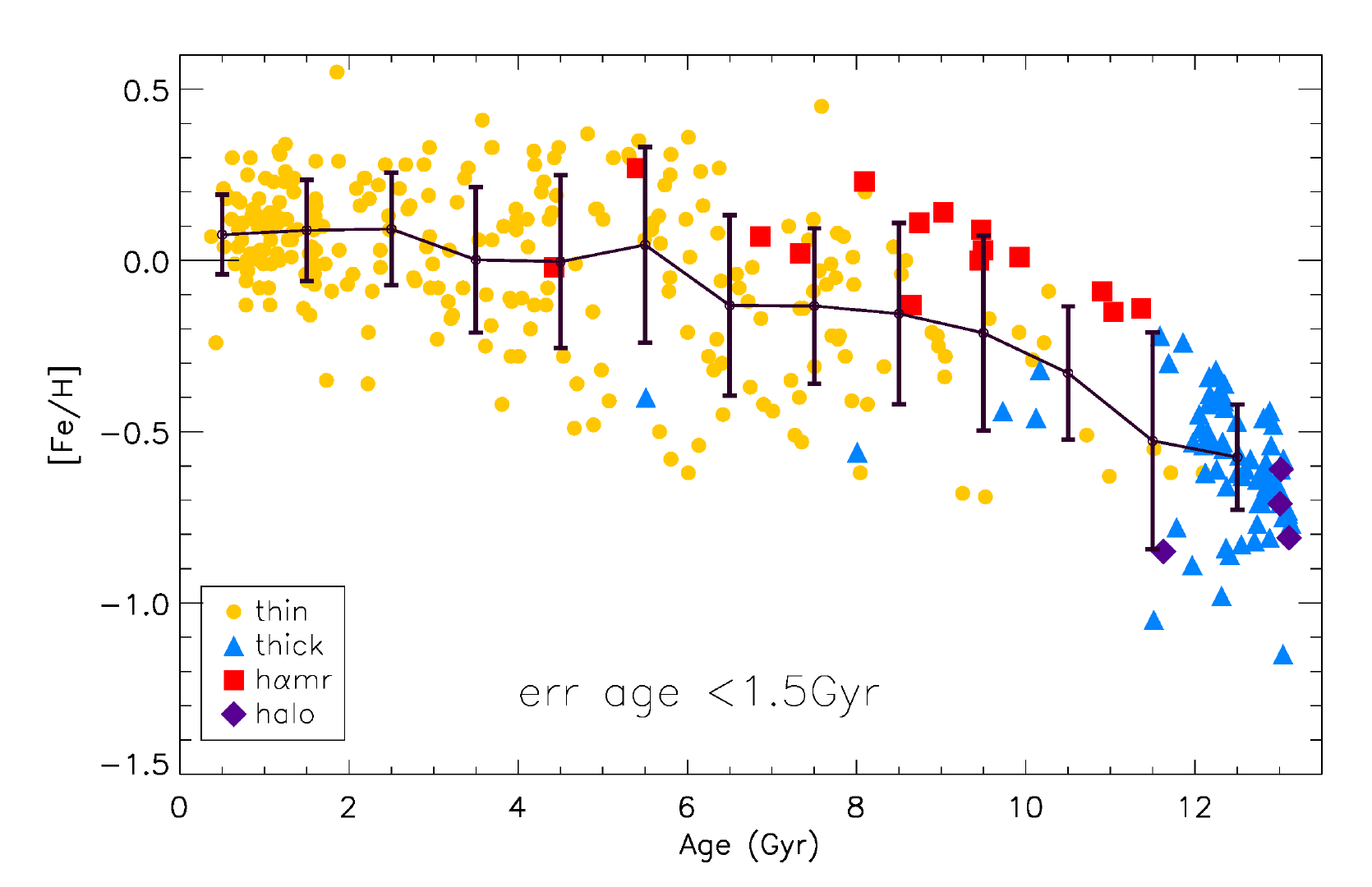}
\includegraphics[width=1\linewidth]{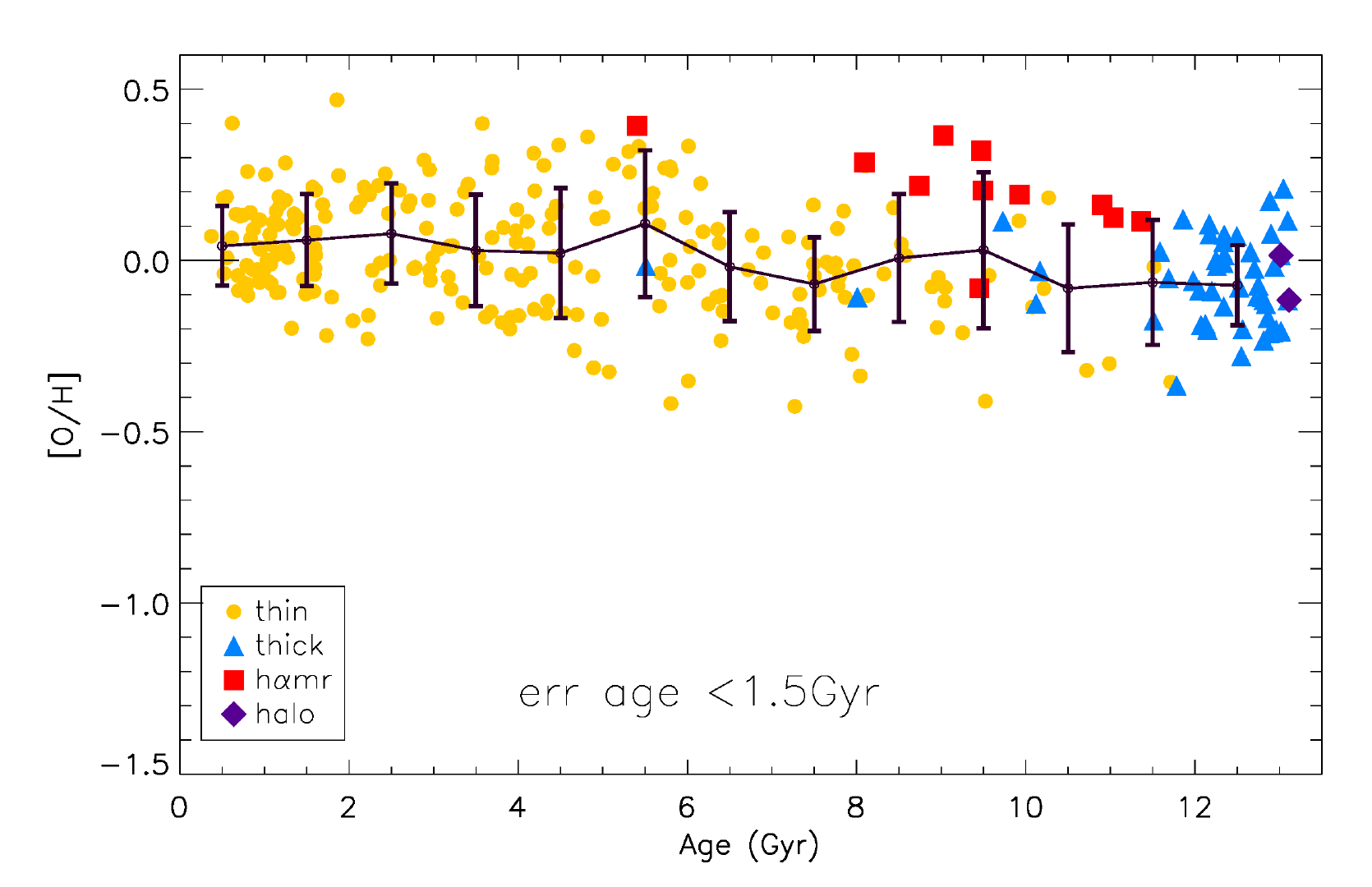}
\caption{[Fe/H] as a function of ages from \textit{Gaia} DR2 for the full sample (upper panel) and for the subsample of stars with errors in age lower than 1.5\,Gyr. The different stellar populations are depicted with different colors and symbols as explained in the legend. The average and standard deviation of [Fe/H] in 1 Gyr size bins are shown with a black line and error bars.} 
\label{feh_ages}
\end{figure}

As a comparison, we also derived masses and ages using parallaxes from the Hipparcos new reduction \citep{vanLeeuwen07} with the same aforementioned method. Hipparcos provides parallaxes for 1051 out of the 1059 stars within our sample. The third set of ages and masses was derived by using the \textit{q$^{2}$} Python package\footnote{https://github.com/astroChasqui/q2} \citep{ramirez14} that considers Yonsei-Yale isochrones \citep{kim02} and spectroscopic parameters (\teff, \logg\ and [Fe/H]). We note that for this set of ages we have used the corrected values of \logg\ presented in DM17. In Fig. \ref{comp_ages} we compare the results we obtain with the three previously mentioned methods. The masses obtained with different methods and/or parallaxes agree very well and the differences are quite small ($<$M$_{q2}$ -- M$_{Gaia}$ $>$ = 0.003$\pm$0.051\,M$_{\odot}$ and $<$M$_{HIP}$ -- M$_{Gaia}$ $>$ = --0.004$\pm$0.030\,M$_{\odot}$) although the comparison between \textit{Gaia} masses and \textit{q$^{2}$} masses show some oscillations around the mean.

However, the situation is different for the ages for which we can find very large differences for some stars although the average differences are not very large: ($<$Age$_{q2}$ - Age$_{Gaia}$ $>$ = 1.12$\pm$1.96\,Gyr and $<$Age$_{HIP}$ -- Age$_{Gaia}$ $>$ = --0.46$\pm$1.28\,Gyr). For example, there are many stars clustered around 8-10\,Gyr in the \textit{q$^{2}$} ages dataset. They are very cool dwarfs and have similar stellar parameters (\teff, \logg\ and [Fe/H]) so the code delivers very similar ages. However, when using parallaxes and magnitudes the degeneracy is broken and their age range increases. Nevertheless, although the differences in age between the three methods present a moderate dispersion there are no large systematic differences. It is also clear that there is a large group of stars whose ages from \textit{Gaia} are much larger than with Hipparcos or \textit{q$^{2}$} (see upper panel of Fig. \ref{comp_ages}). Most of the stars that have an age with \textit{Gaia} greater than with Hipparcos also have smaller parallaxes (see Fig. \ref{comp_ages_plx}) and the difference in age is not correlated with the magnitude.

Several works have determined the systematic offset between \textit{Gaia} DR2 parallaxes and other samples with independently derived parallaxes, all of them finding that \textit{Gaia} parallaxes are lower by 0.03-0.09 mas. However, such offset becomes more important for more distant stars. For example, in \citet{stassun18} the comparison between eclipsing binaries parallaxes and \textit{Gaia} DR2 shows that they are almost the same down to 4 mas. Since in our sample only three stars have a parallax smaller than 4 mas we decided to add a conservative value of 0.03 mas as suggested by the \textit{Gaia} collaboration \citep{lindegren18_gaia}. Moreover, we also increased the errors in parallaxes to consider the $\sim$\,30\% underestimation in uncertainties for bright stars \citep{luri18_gaia,arenou18_gaia}. However, we remark that the increase of errors by such amount hardly affects the derived ages. We note that the average parallax difference for the 1047 stars in common between \textit{Gaia} DR2 and Hipparcos is --0.239\,$\pm$\,1.206\,mas, greater than the comparison of the full sample made by \citet[][--0.118\,$\pm$\,0.003\,mas]{arenou18_gaia}. In any case, we decided to use ages derived with \textit{Gaia} DR2 parallaxes as final ages (given that those parallaxes are much more precise than Hipparcos). In order to have more reliable results we will consider in next subsections the 354 stars with errors in age lower than 1.5\,Gyr (265 thin disk stars, 15 h$\alpha$mr stars, 70 thick disk stars and 4 halo stars). These 354 stars have a large range in parameters \teff: 5010$-$7212K (95\% between 5300$-$6500\,K), \logg: 3.73$-$4.71\,dex (92\% between 4.0$-$4.6\,dex), [Fe/H]: --1.15$-$0.55\,dex (93\% between --0.7$-$0.4\,dex). This chosen limit in age error is a compromise between having a reliable set of stellar ages but still with a sufficient number of stars for a meaningful analysis. However, we note that by cutting the age error at 1 or 2\,Gyr instead of 1.5\,Gyr the final conclusions of this work would not change. The final ages and masses are listed in Table \ref{tab:ages}.

\section{Temporal evolution of [Fe/H] and [$\alpha$/Fe]}\label{sec:feh_age}

In the upper panel of Fig. \ref{feh_ages} we depict the metallicity-age relation in the full sample. There is an overdensity of stars around 5-6\,Gyr caused by most of the cool stars ($\lesssim$\,5100\,K) for which the errors in age are typically above 4\,Gyr. Therefore, we show in the lower panel of the same figure the subsample with more precise ages (error in age lower than 1.5\,Gyr). It is clear from the figure the large dispersion of ages at a given metallicity as previously observed in the solar neighborhood \citep[e.g.,][]{haywood13,bensby14,bergemann14} or in other specific samples such as the dwarf stars in GALAH \citep{buder18} or the giant stars in Kepler \citep{silva-aguirre18}, LAMOST-Kepler \citep{wu18} or APOGEE \citep{feuillet18}. This confirms the weak age-[Fe/H] correlation first pointed out by \citet{edvardsson93} which is assumed to be caused by radial migration \citep{sellwood02}. In the absence of radial migration we would expect to have a stronger age-[Fe/H] correlation. Furthermore, the dispersion in metallicity increases with age being 0.13\,dex for stars younger than 2\,Gyr, 0.19\,dex between 2 and 4\,Gyr and around 0.26\,dex for older stars. However, there is a decrease of the dispersion to 0.16\,dex for stars older than 12\,Gyr (i.e., thick disk stars) which might be caused by the low number of metal-poor stars in our sample. Recent observations seem to suggest no strong radial metallicity gradient for the thick disk \citep{cheng12,hayden15,allende06}. This would mean that either there was not a radial metallicity gradient at the formation of the thick disk (meaning that the stars were formed from a well mixed material) or that there was a gradient that was flattened out by efficient radial migration. However, efficient radial migration in the thick disk would make the metallicity dispersion wider \citep[e.g.,][]{minchev13} which, if real, contradicts our observed trend. Therefore, our observations together with a current flat metallicity gradient suggests that the thick disk was probably formed from a well mixed material \citep{haywood15} and with no radial metallicity gradient.

The average [Fe/H] per age bin keeps around 0.08\,dex up to $\sim$\,3\,Gyr, then it decreases to solar value for stars up to $\sim$6\,Gyr. At ages\,$\gtrsim$\,6\,Gyr, however, there is a more clear decreasing trend of [Fe/H] with age, reaching average values of -0.57\,dex (12-13\,Gyr) and -0.75\,dex ($>$\,13\.Gyr). The most metallic star (HD\,108063, [Fe/H]=0.55) has an age of 1.81\,$\pm$\,0.06\,Gyr but the majority of most metal-rich (with [Fe/H]$\sim$0.4\,dex) stars are indeed close to 4\,Gyr. It is very probable that the most metal-rich stars were formed in the inner Galaxy and because they are older, they had time to radially migrate into the solar neighborhood \citep[e.g.,][]{sellwood02}. Indeed, a recent study by \citet{minchev18} using our sample shows that the most metal rich stars have the smallest birth radii.

If we consider only those stars with a low error on age ($<$\,1.5\,Gyr) (middle panel of Fig. \ref{feh_ages}) we can see that only a few thick disk stars have ages lower than 11\,Gyr and \textit{h$\alpha$mr} stars are always older than 4\,Gyr. On the other hand very few thin disk stars are older than 10\,Gyr in general agreement with previous works \citep[e.g.,][]{haywood13} but in contrast with the recent works by \citet{wu18} and \citet{silva-aguirre18}, where ages have been derived using asteroseismic constraints. We can also see a certain pile up of stars around 13.5\,Gyr. This is an artifact of the prior we input as maximum age to avoid having ages older than the age of the Universe. 

In the lower panel of Fig. \ref{feh_ages} the relation of [O/H] is shown. We have taken the oxygen abundances from \citet{bertrandelis15}, derived for the HARPS-GTO sample. We note that those abundances were calculated with the previous set of parameters so we have rederived them here using the corrected \logg\ from DM17\footnote{Since oxygen is only derived for stars hotter than 5200\,K the updated linelist used to derive stellar parameters for cool stars would not be applied in this case.} although we note that the changes are not substantial. Oxygen is considered a "pure" $\alpha$ element, mostly produced by massive stars and ejected in SNeII \citep[e.g.,][]{woosley95,andrews17} and its production along the time is quite constant in comparison with iron, with a maximum dispersion of 0.2\,dex, around 6\,Gyr, and average [O/H] values between --0.1 and 0.1\,dex across all the ages.

\begin{figure}
\centering
\includegraphics[width=1\linewidth]{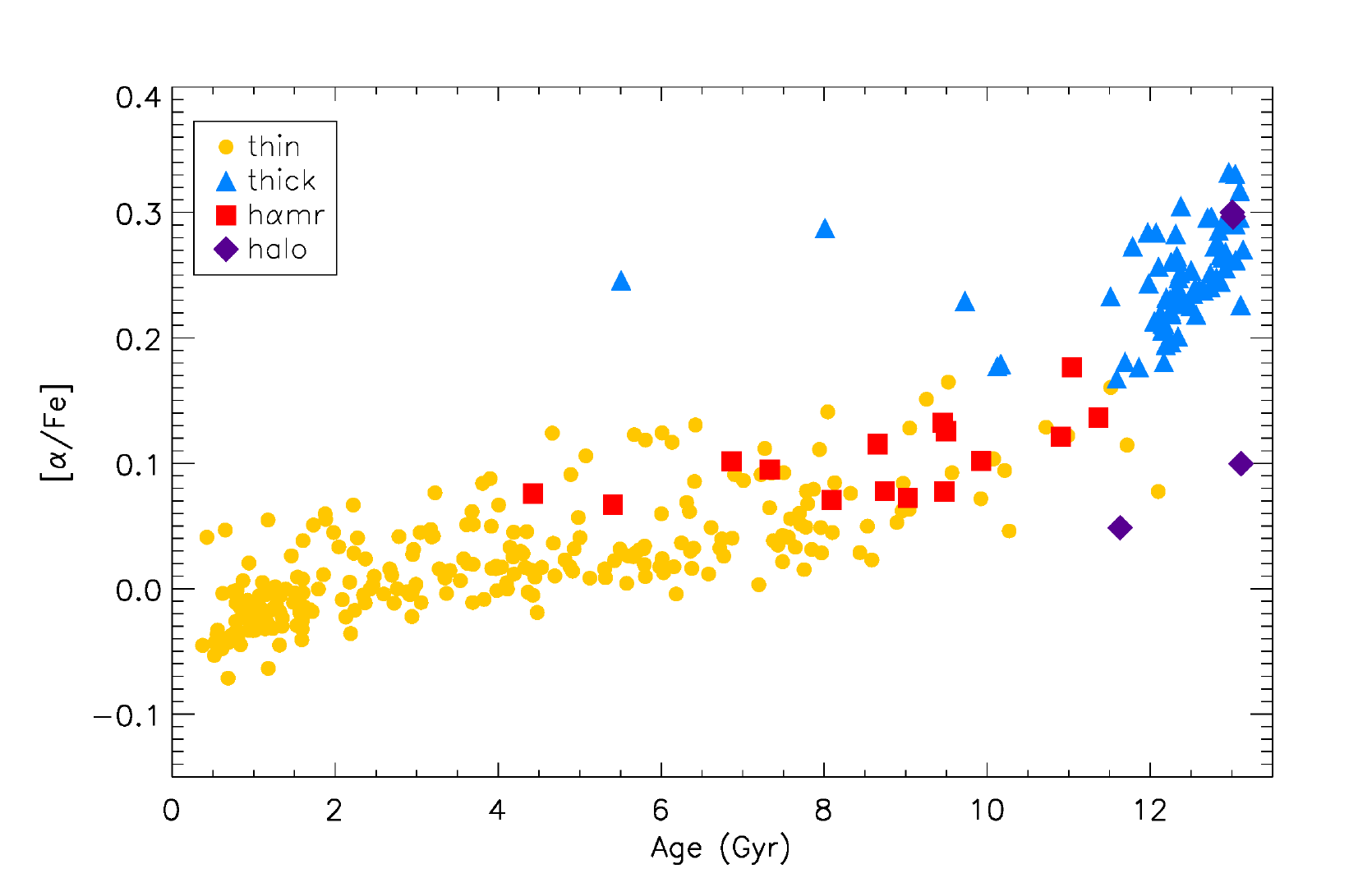}
\includegraphics[width=1\linewidth]{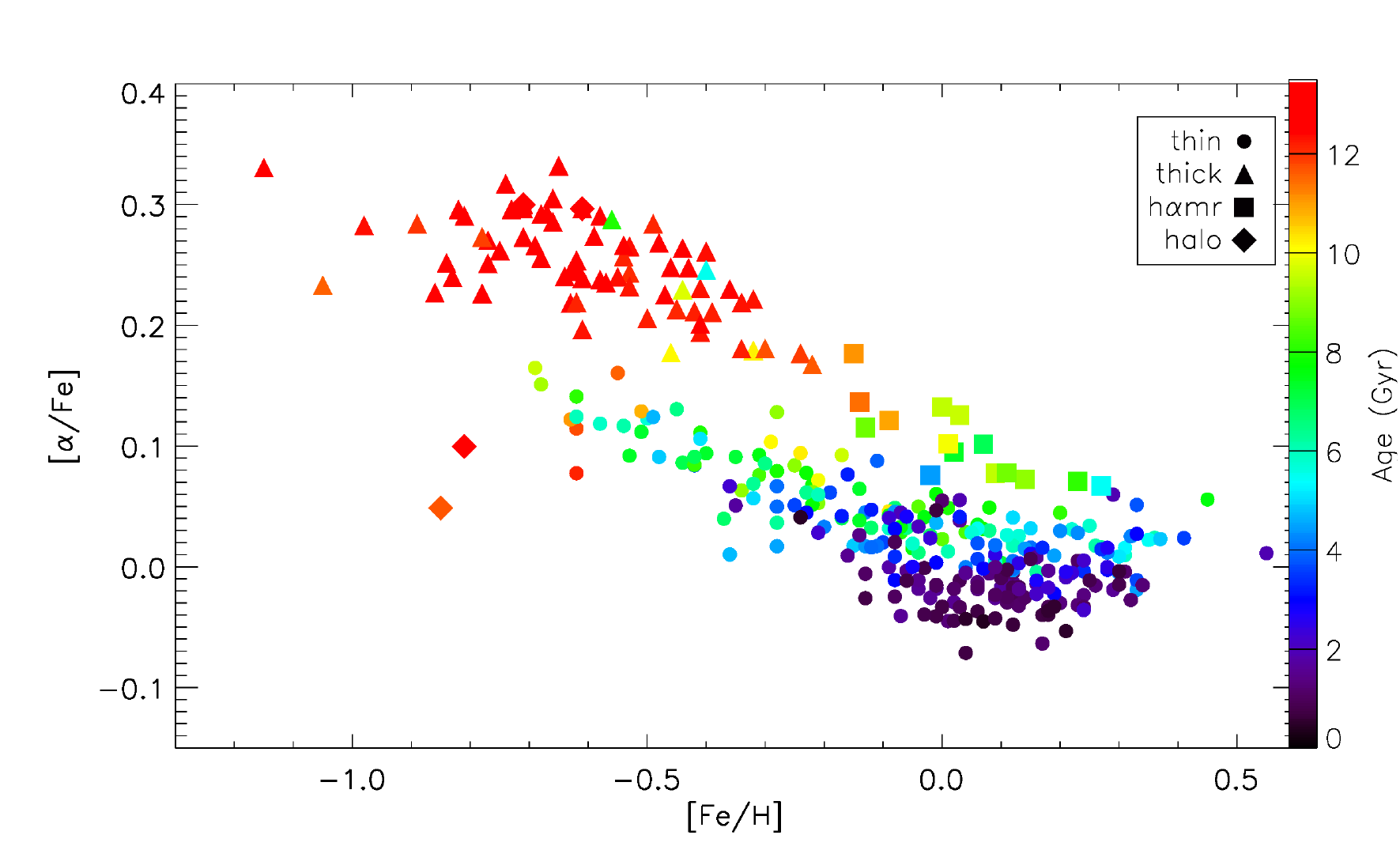}
\caption{[$\alpha$/Fe] as a function of [Fe/H] for the subsample of stars with errors in age from \textit{Gaia} DR2 lower than 1.5\,Gyr. The different stellar populations are depicted with different colors and symbols as explained in the legend.} 
\label{alpha_feh_ages}
\end{figure}

Several works have proposed that [$\alpha$/Fe] is a better indicator of age than [Fe/H] and this is also confirmed by our results (see the upper panel of Fig. \ref{alpha_feh_ages} were $\alpha$ is the average of Mg, Si, and Ti). In contrast with metallicity, the dispersion of [$\alpha$/Fe] ratios is much lower, with values of 0.03\,dex up to 4\,Gyr, 0.04\,dex from 4 to 8\,Gyr, 0.06\,dex from 8 to 12\,Gyr and 0.04\,dex for stars older than 12\,Gyr. The abundances of $\alpha$ elements with respect to iron in thin disk and \textit{h$\alpha$mr} stars show a clear increasing linear trend with age (with Spearman correlation coefficients, $\rho$, of 0.78 and 0.72, respectively). The tail of the distribution at old ages, formed by thick disk stars, also presents an increasing trend, though with a somewhat lower $\rho$ of 0.59. This lower significance is mostly produced by the shorter age range of our chemically defined thick disk stars. In other works there is not distinction between thick disk and \textit{h$\alpha$mr} stars, thus having thick disk stars as young as 8-9 Gyr \citep[e.g.,][]{haywood13,bensby14}. If we consider the \textit{h$\alpha$mr} stars as part of the thick disk, the Spearman correlation coefficient for the [$\alpha$/Fe]-age relation would be 0.74. Moreover, the lower significance of the correlation is also partly caused by the existence of a few "young" stars ($\sim$5-10\,Gyr) with a similar $\alpha$ content as their older counterparts. In contrast, this region of intermediate ages and high-$\alpha$ abundances is well populated in the works by \citet{wu18} and \citet{silva-aguirre18} producing a rather flat behavior with age for the high-$\alpha$ sequence. Several works have found the existence of young $\alpha$-rich stars \citep[e.g.,][]{fuhrmann12,fuhrmann17,haywood15}, which seem to be more common in the inner disk of the Galaxy \citep{chiappini15}. These objects might be evolved blue stragglers \citep[e.g.,][]{fuhrmann11,jofre16} or have been formed near the end of the Galactic bar \citep{chiappini15}. 

Regarding the shape of the [$\alpha$/Fe]-age trend, the work by \citet{haywood13}--that uses our HARPS-GTO sample of abundances--finds a change of slope around 8\,Gyr, at which older stars show a steeper trend of [$\alpha$/Fe] with age. In that work the oldest thin disk stars are about 10\,Gyr, similar to the present study. However, in the Kepler and LAMOST-Kepler samples there are a significant number of older thin disk stars that follow the trend of younger stars and thus avoid the change of slope found in \citet{haywood13} and also found in the present study but at an older age. Nevertheless, we note here that the trends found by \citet{wu18} and \citet{silva-aguirre18} might be difficult to compare with those in the present study or in the work by \citet{haywood13} due to the much larger errors in age for old stars in those two samples of red giants. 

\begin{figure}
\centering
\includegraphics[width=1\linewidth]{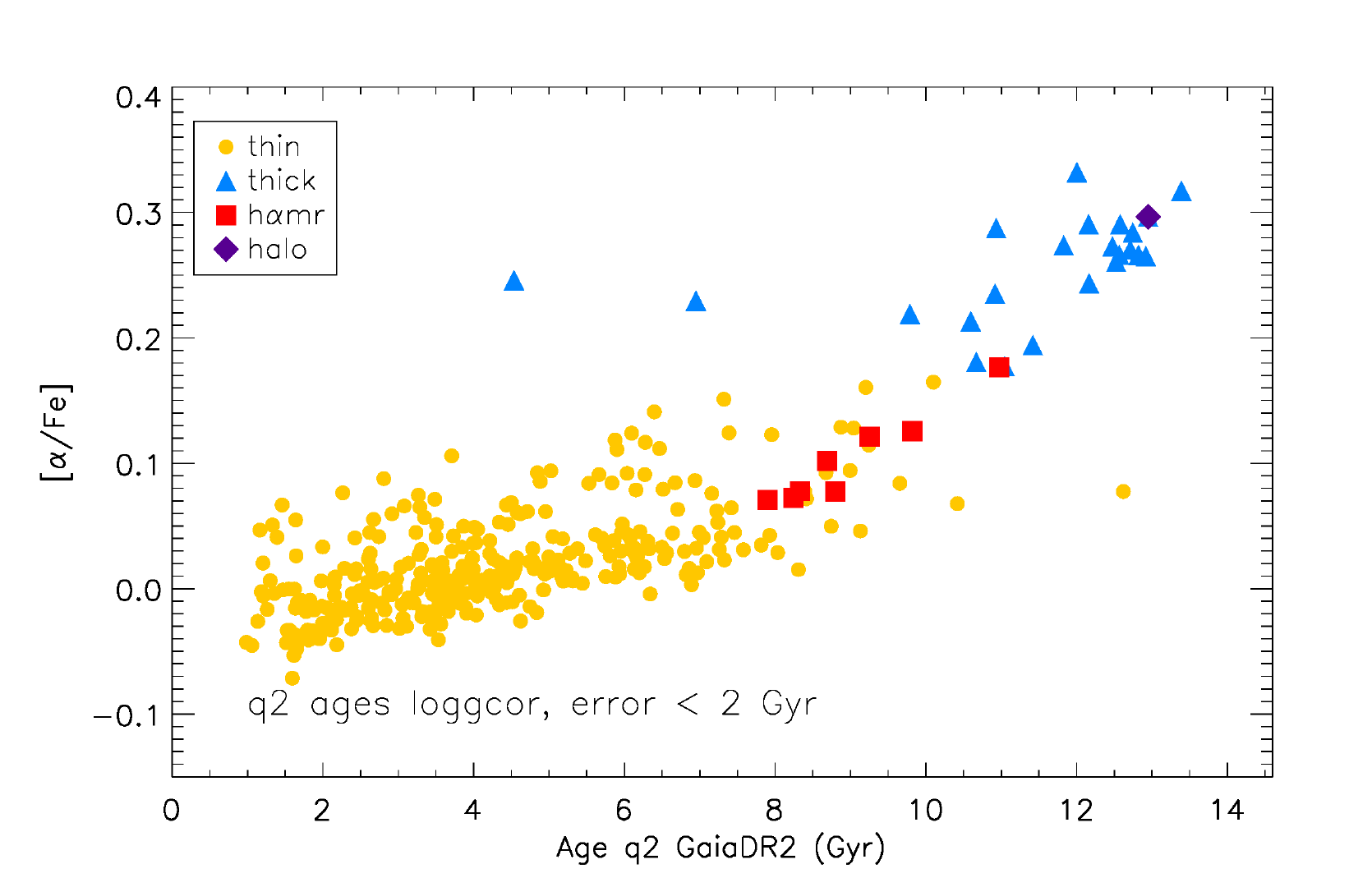}
\includegraphics[width=1\linewidth]{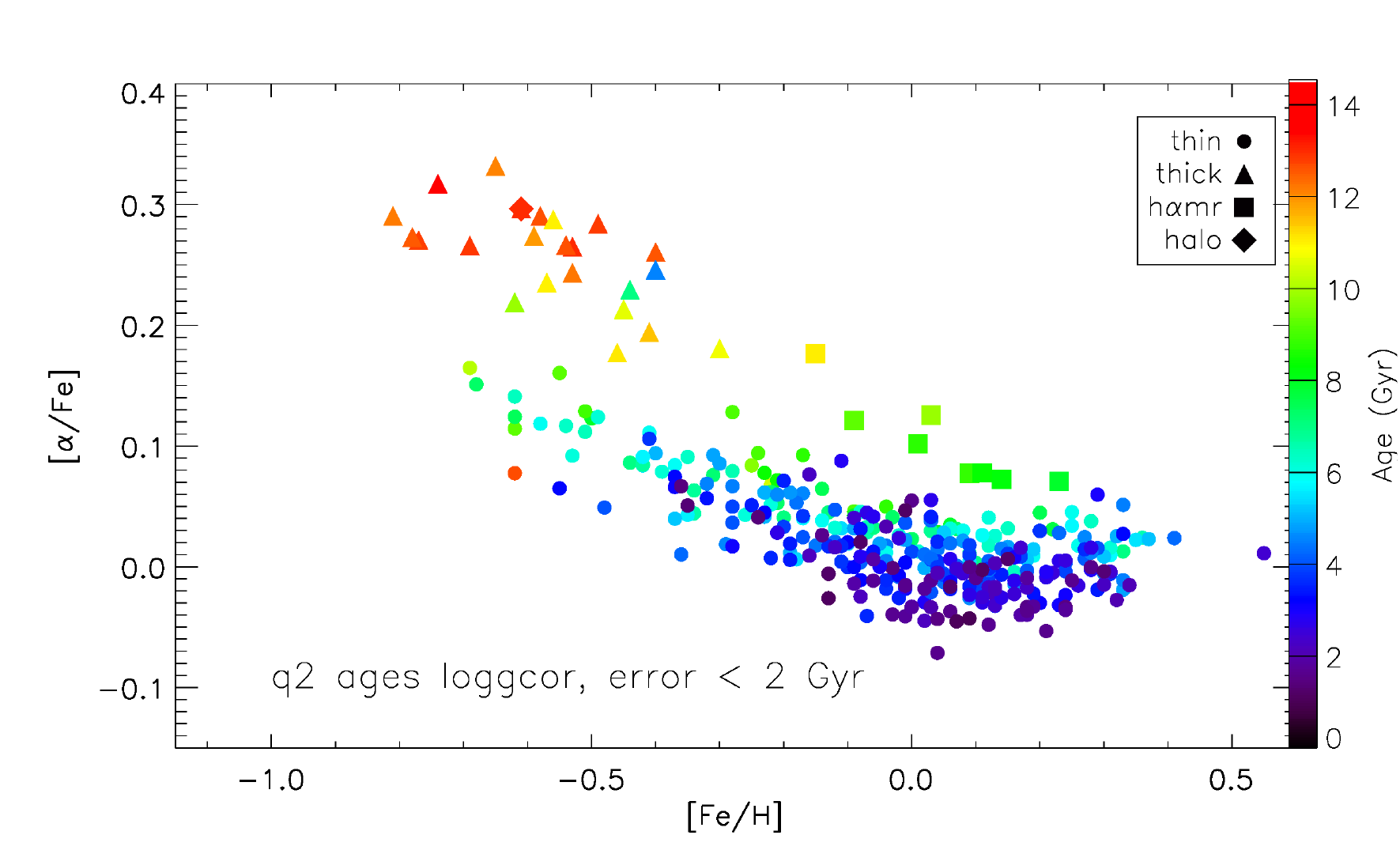}
\caption{[$\alpha$/Fe] as a function of [Fe/H] for the subsample of stars with errors in age from \textit{q$^{2}$} lower than 2\,Gyr. The different stellar populations are depicted with different colors and symbols as explained in the legend.} 
\label{alpha_feh_ages_q2}
\end{figure}

In Fig. \ref{alpha_feh_ages_q2} we show the same relations for [$\alpha$/Fe] using the ages from \textit{q$^{2}$}, that are derived with the same isochrones as in \citet{haywood13} and making use of \logg\ instead of parallaxes\footnote{For this figure with \textit{q$^{2}$} ages we selected the stars with errors in age lower than 2\,Gyr instead of 1.5\,Gyr in order to have a reasonable amount of stars.}. The trends are quite similar as those using ages derived with \textit{Gaia} parallaxes. However, there are hardly no thin disk stars above 10 Gyr and [$\alpha$/Fe] increases more steeply with age after $\sim$\,8\,Gyr whereas when using \textit{Gaia} ages the change of slope seems to happen around 10-11\,Gyr. In the dwarfs sample within GALAH presented by \citet{buder18} there are also many stars older than 11\,Gyr with low [$\alpha$/Fe] as in the works by \citet{wu18} and \citet{silva-aguirre18}. Nevertheless, those ages might not be very precise due their large errors as reported by \citet{buder18}. However, the [$\alpha$/Fe] ratios by \citet{buder18} seem to present a steep trend with age for older ages (still with a high dispersion) in contrast with the work by \citet{wu18} and \citet{silva-aguirre18}. Whether the steeper increase of [$\alpha$/Fe] with age for older stars\footnote{Here we refer to older thick or thin disk stars since for halo stars the relation might be different. For example, \citet{fernandez_alvar15} reported that the outer halo (presumably younger) shows generally lower [$\alpha$/Fe] ratios than the inner (presumably older) halo.} is taking place only in the solar neighborhood or if it is extendable to other parts of the Galaxy will be better understood with future releases of large spectroscopic surveys. In addition, the use of different methods to derive ages (e.g., asteroseismic constrains, using \logg\ instead of parallax, applying a Bayesian approach instead of a $\chi^{2}$ fitting) or the use of different sets of isochrones might produce different trends as shown in Figs. \ref{alpha_feh_ages} and \ref{alpha_feh_ages_q2}. The works by \cite{haywood13} and \cite{bensby14} both use Yonsei-Yale isochrones and they find the change of slope in the [$\alpha$/Fe]-age relation around 8-9\,Gyr in a similar way as our sample with \textit{q$^{2}$} ages (which also uses Yonsei-Yale). However, our ages obtained with PARSEC isochrones tend to be older which might be the reason for the change of slope occurring $\sim$2\,Gyr later and the fact of having a handful of thin disk stars older than 10\,Gyr. In a similar way, the works by \citet{anders18,minchev18} that derive ages for our sample using the PARSEC isochrones within the StarHorse package, find a change of the [$\alpha$/Fe]-age slope at $\sim$10\,Gyr. On the other hand, the work by \citet{buder18} uses the Dartmouth isochrones which also produce larger ages than Yonsei-Yale isochrones \citep[see Fig. 3 in][]{haywood13}. This might explain why that work finds so many thin disk stars older than 11\,Gyr and why the change of slope of the [$\alpha$/Fe]-age seems to happen at a bit older age than that of \cite{bensby14} \citep[although still compatible within the uncertainties, see Fig. 14 in][]{buder18}.

The correlation between [$\alpha$/Fe] and [Fe/H] is depicted in the lower panel of Fig. \ref{alpha_feh_ages} with a color scale for the ages. Most of the stars with [$\alpha$/Fe]\,$\gtrsim$\,0.2\,dex (92\%) have ages greater than 10\,Gyr and belong to the thick disk. This fact shows that the formation of the thick disk was fast and took place mostly before the thin disk. On the other hand, the thin disk stars show a great dispersion in ages, which increase as we move to lower [Fe/H] when considering stars of a similar [$\alpha$/Fe]. If we look at the oldest thin disk stars, they have a similar age and [$\alpha$/Fe] (around 0.1\,dex) as the oldest \textit{h$\alpha$mr} (orange-green squares) but they are more metal-poor. \citet{haywood13} proposed that this group of older thin disk stars must have been formed in a different part (specifically the outer disk) than the \textit{h$\alpha$mr} stars\footnote{The authors of that work consider the \textit{h$\alpha$mr} to belong mostly to the thin disk sequence.} since they have different metallicities and a higher rotational component in their velocities. The study of the AMBRE:HARPS sample by \citet{minchev18} also shows that those metal-poor thin disk stars have the largest birth radii. We can also see that there is a very clear stratification in ages (older as [$\alpha$/Fe] increases) when looking at stars at a given [Fe/H], running from $<$\,2\,Gyr for the most $\alpha$-poor (at [Fe/H]\,$\gtrsim$\,--0.2\,dex) to 8-10\,Gyr as we reach the \textit{h$\alpha$mr} stars. Therefore, \textit{h$\alpha$mr} stars are clearly separated from the metal-rich thin disk both in terms of [$\alpha$/Fe] and age. A similar conclusion is also reached with the analysis by \citet{anders18}. Since most of our thick disk stars have ages above 12 Gyr it is difficult to see a temporal evolution in the thick disk as metallicity diminishes. However, in the lower panel of Fig. \ref{alpha_feh_ages_q2} where the same plot is done using the \textit{q$^{2}$} ages, thick disk stars tend to be older as [$\alpha$/Fe] increases and [Fe/H] decreases. Nevertheless we note that the number of thick disk stars with a low error in age is significantly reduced in this set of ages. We note here that this temporal evolution in the thick disk would be more obvious if we were to consider the \textit{h$\alpha$mr} stars (or at least the oldest ones) as part of the thick disk. For example, in the works by \citet{bensby14,haywood13} the metallicity of the thick disk can reach values above solar.

\begin{figure*}
\centering
\includegraphics[width=19cm]{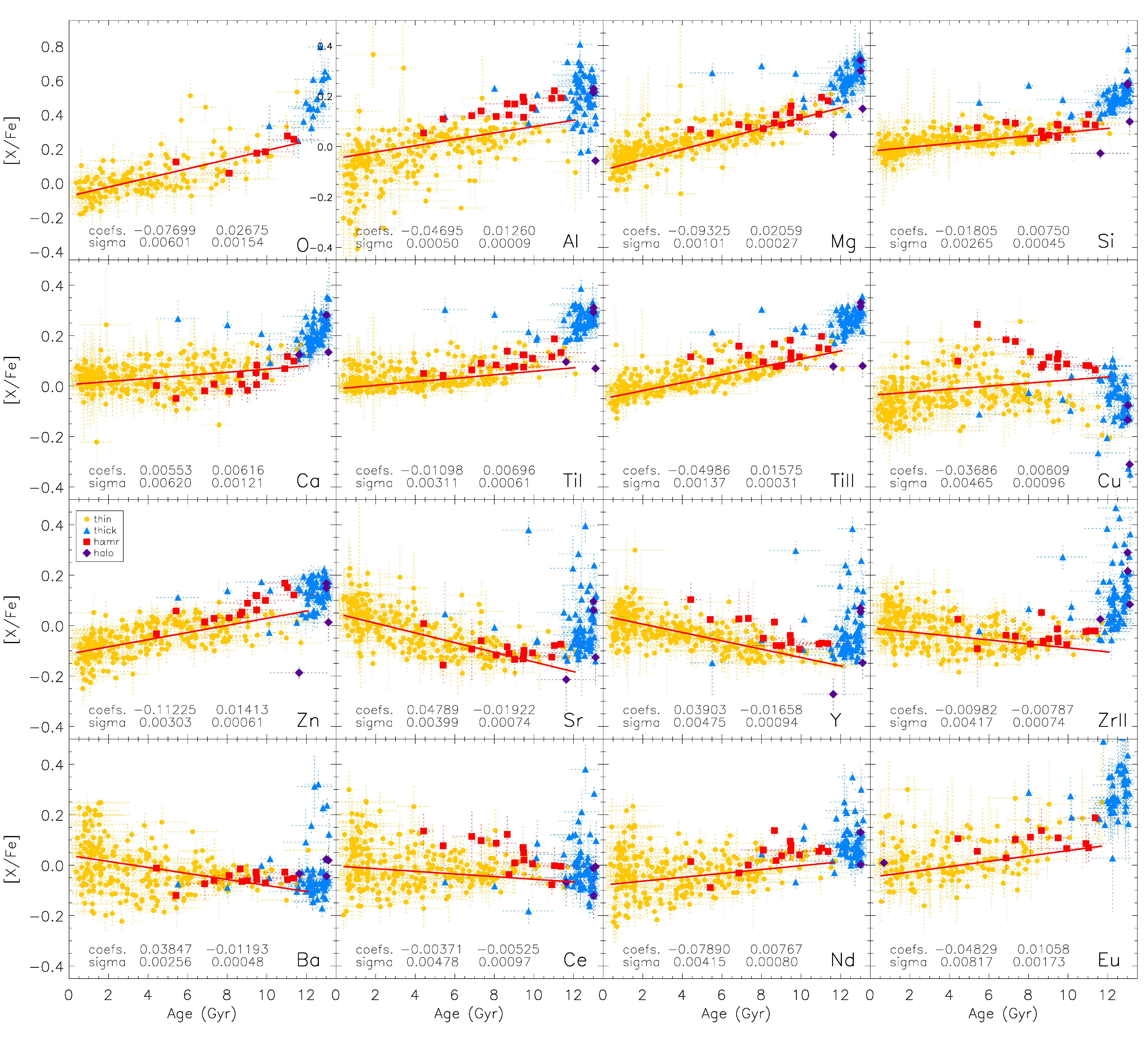}
\caption{[X/Fe] as a function of age for stars with an error in age smaller than 1.5 Gyr. The different stellar populations are depicted with different colors and symbols as explained in the legend. We note the different size of y axis for oxygen with respect to the rest of elements. The red line is a weighted linear fit to the thin disk stars to guide the eye on the general behavior of the trends. The \textit{coefs.} values in each panel are the abscissa origin and the slope of the fit, respectively, together with the error (\textit{sigma}) of each coefficient.} 
\label{all_age}
\end{figure*}

\begin{figure*}
\centering
\includegraphics[width=19cm]{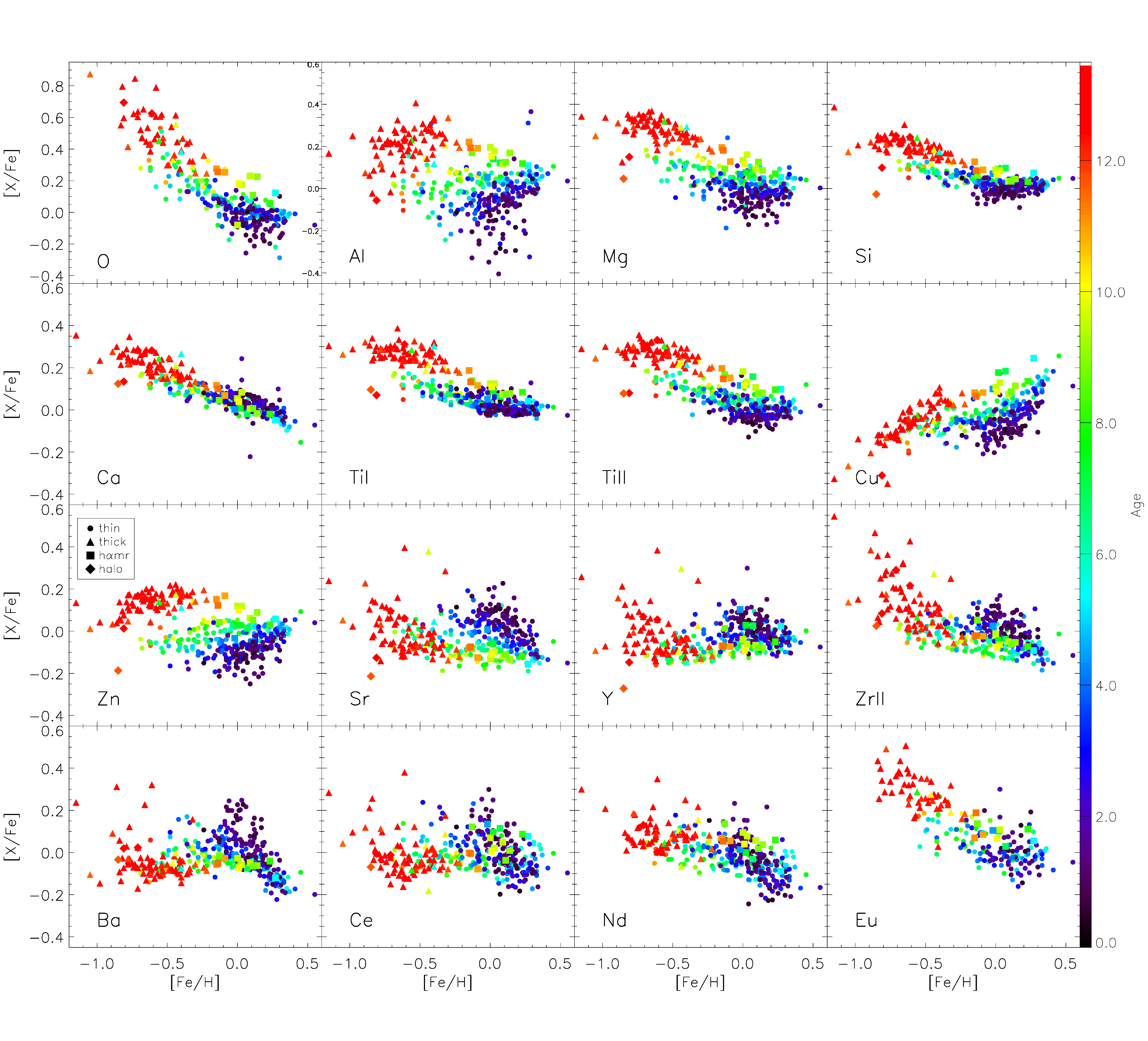}
\caption{[X/Fe] as a function of [Fe/H] for stars with an error in age smaller than 1.5 Gyr. We note the different size of y axis for oxygen with respect to the rest of elements. The circles, triangles, squares and diamonds are the stars from the thin disk, thick disk, \textit{h$\alpha$mr} and halo.} 
\label{all_feh_color_age}
\end{figure*}

\begin{figure*}
\centering
\includegraphics[width=19cm]{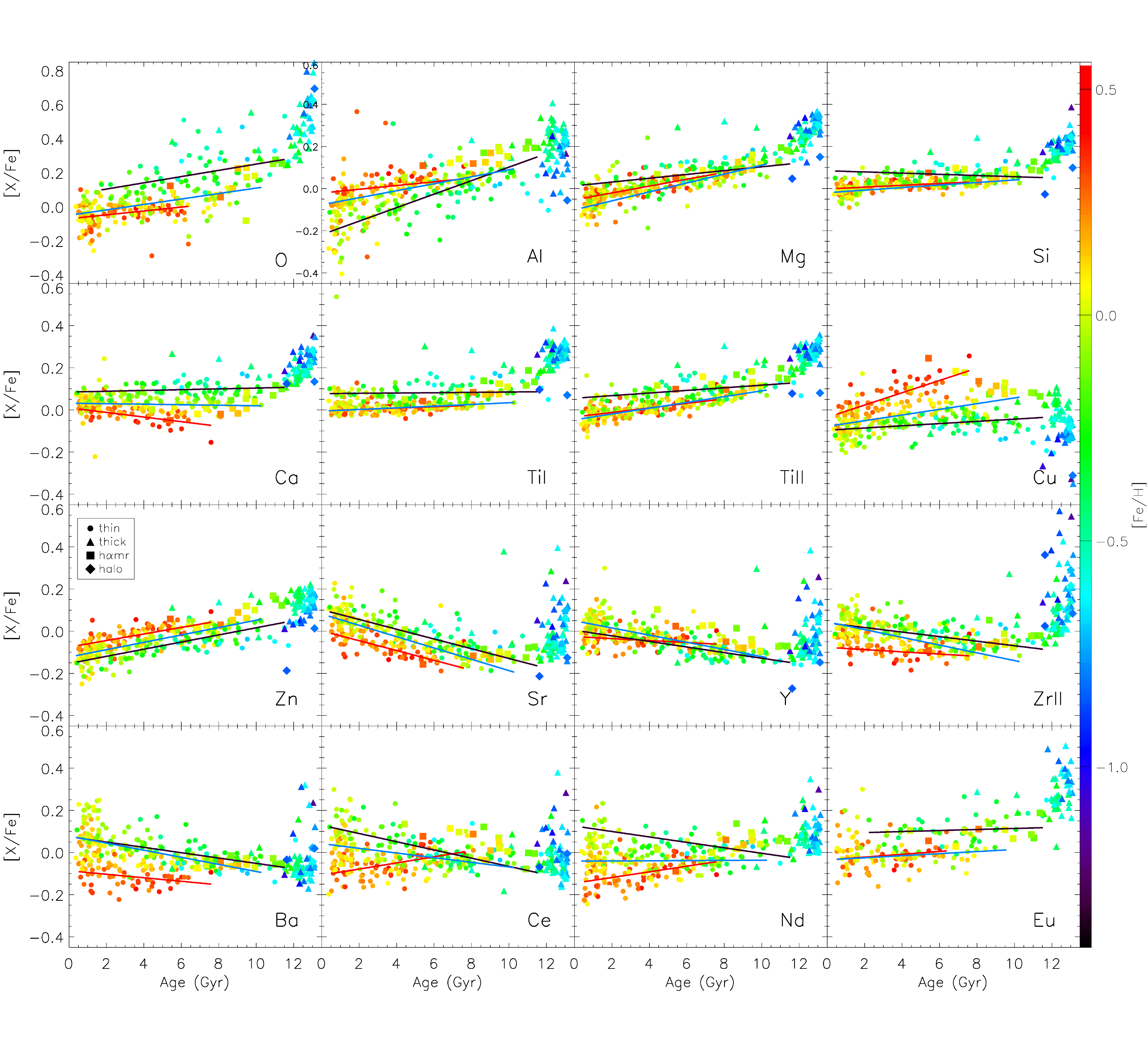}
\caption{[X/Fe] as a function of age for stars with an error in age smaller than 1.5 Gyr. We note the different size of y axis for oxygen with respect to the rest of elements. The red, blue, and black lines are weighted linear fits to thin disk stars with [Fe/H]\,$>$\,0.2\,dex, --0.2\,$<$\,[Fe/H]\,$<$\,0.2\,dex and --0.6\,$<$\,[Fe/H]\,$<$\,--0.2\,dex, respectively.} 
\label{all_age_color_feh}
\end{figure*}

\begin{figure*}
\centering
\includegraphics[width=1\linewidth]{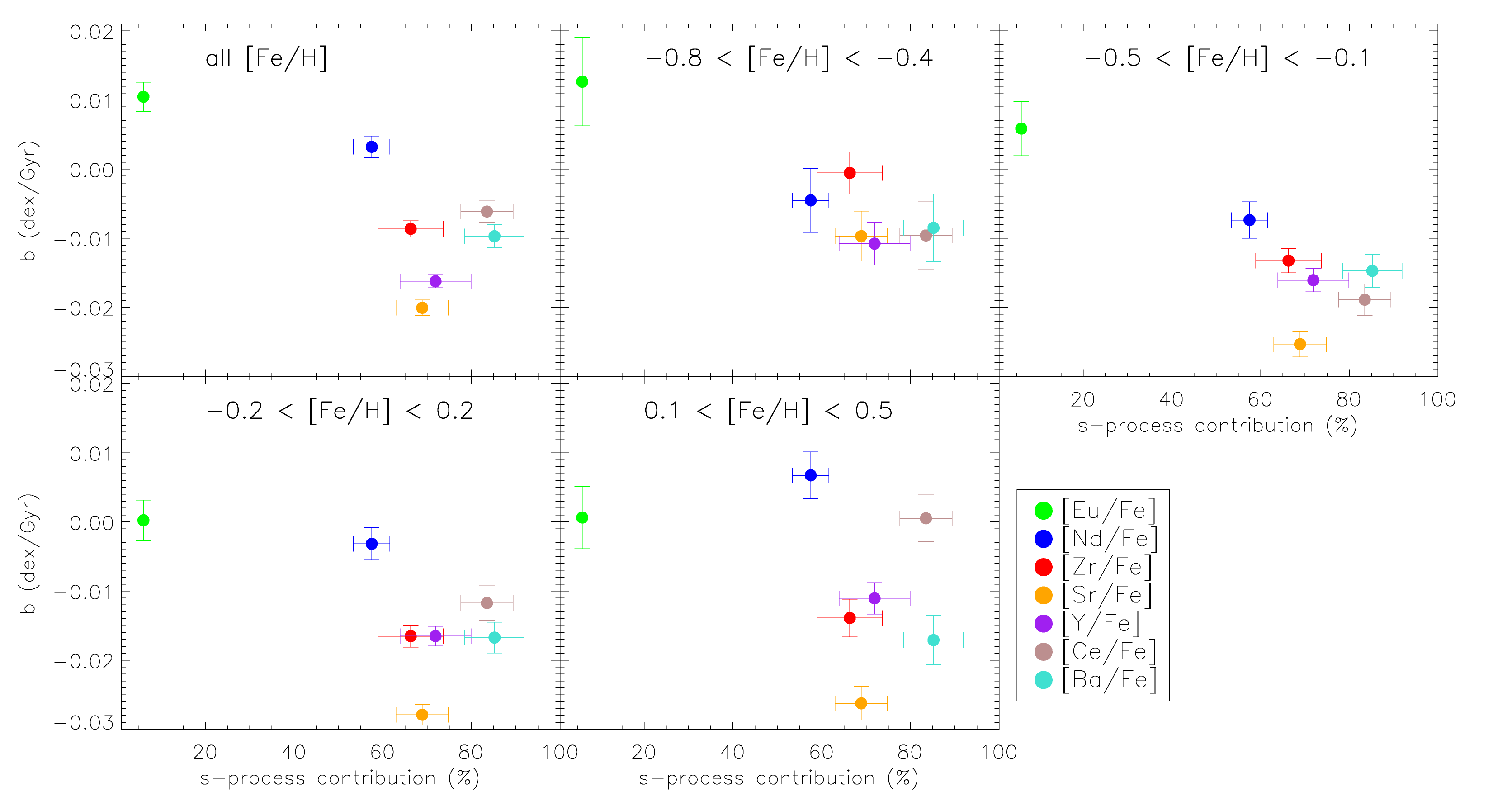}
\caption{Slopes (dex/Gyr) of the [X/Fe]-age relations ([X/Fe]=a+b$\cdot$Age) for neutron capture elements as a function of \textit{s}-process contribution for thin disk stars att all metallicities and in four metallicity bins with a width of 0.4\,dex} 
\label{heavy_slopes}
\end{figure*}

\section{Dependence of individual [X/Fe] ratios on [Fe/H] and age}\label{sec:xfe_age}

In Fig. \ref{all_age} we show the [X/Fe] ratios of the elements presented in DM17 as a function of age for the different populations in our sample together with a linear fit to thin disk stars. To complement the qualitative study of $\alpha$ elements we have added the rederived oxygen abundances from \citet{bertrandelis15}, as shown in Fig. \ref{feh_ages}. A first look at this figure allows us to see the expected general trends. The ratios of $\alpha$ elements O, Mg, Si, Ca, and Ti respect to Fe show an increasing trend toward older ages with O and Mg showing the steepest trends (the slopes for thin disk stars are 0.026 and 0.02 dex/Gyr, respectively). These trends are in general similar to those observed in other works such as \citet{nissen15,bedell18,anders18,feuillet18,buder18}. Those elements are mostly produced in SNeII meanwhile Fe is produced mainly by SNeIa. Since the progenitors of SNeII are more massive than the progenitors of SNeIa, the ratios [$\alpha$/Fe] will be higher at early ages in the Galaxy because massive stars have shorter lifetimes. However, the contribution of SNeII to O and Mg is higher than to the other $\alpha$ elements, which are also partially produced by SNIa \citep[e.g.,][]{nomoto13}. Therefore, that could explain the steeper age trend for O and Mg. Since Al is mainly produced by core-collapse SNe \citep[e.g.,][]{andrews17}, the abundance-age trend is also similar to that of $\alpha$ elements, with a steep slope but a higher dispersion. However, we note that this positive trend might be limited by the lower limit in [Fe/H] of our sample because stars with [Fe/H]\,$\lesssim$\,--1.5\,dex (hence, old) show quite low [Al/Fe] values \citep[see e.g., the compilation made by][]{prantzos18}. Finally, the element Zn also increases with age since it has an important contribution by neutrino winds during supernovae explosions of massive stars \citep[e.g.,][and references therein]{bisterzo05}. Interestingly, this is the only element to show a similar behavior with age for both thin and thick stars and might represent a better global age proxy than [$\alpha$/Fe] ratios.

It is also clear from Fig. \ref{all_age} that thick disk stars present a stronger enrichment in $\alpha$ elements when compared to thin disk stars at similar age. This extra enrichment in thick disk stars is also observed for the \textit{r}-process element Eu, which is mainly produced by neutron star mergers \citep{drout17,cote18} and core-collapse supernovae \citep{travaglio99}. Both elements have massive progenitors, and in turn [Eu/Fe] also shows a rise toward older ages. Therefore, these trends support the results by \citet{snaith15} showing that the star formation rate was more intense in the thick disk than in the thin disk. 

On the other hand, \textit{s}-process elements show a decreasing trend of [X/Fe] as age increases. These elements are mainly produced in low-mass AGB stars so we can expect them to increase with time (for younger stars) due to the increasing and delayed contribution of low-mass stars as the Galaxy evolves. However, there seems to be a change of slope around 8\,Gyr caused by thick disk stars, similar to the results by \citet{battistini16} for Sr and Zr\footnote{We note that in that work the thin and thick disk stars are separated based on their age, being younger or older than 7 and 9\,Gyr, respectively.}. The results of \citet{spina18} and \citet{magrini18} for Ba and Ce also seem to show a change of slope around 7\,Gyr. We can observe that the abundance ratios of light-\textit{s} elements (Sr, Y, Zr) are on average larger in the thick disk than in the thin disk for similar ages while the heavy-\textit{s} elements (Ba, Ce) present similar enrichments at a given age. This means that the production of light-\textit{s} elements was more efficient than that of heavy-\textit{s} elements in the thick disk. Whether the overproduction of light-\textit{s} elements at lower metallicities is produced by intermediate mass AGB stars or rotating massive stars is still unclear \citep[see the discussion in DM17;][]{bisterzo16,prantzos18}.

In Fig. \ref{all_age} we can also observe that heavy-\textit{s} elements present a great dispersion at young ages whereas Sr and Y present tighter correlations with age. This seems to be caused by the different [X/Fe]-age relations depending on metallicity as explained in next section. Nd is also considered as a heavy-\textit{s} element but it has a remarkable contribution of 44\% from the \textit{r}-process \citep{arlandini99,bisterzo16}, which is produced by massive stars (neutron star mergers or core-collapse supernovae), thus the behavior with age is balanced by the two sources and shows a rather flat trend. This is also the case of Cu, which has an important part produced through the weak-\textit{s} process in massive stars during core He and shell C burning, where neutrons are provided by the $^{22}$Ne($\alpha,n$)$^{25}$Mg reaction. Since the neutron source, $^{22}$Ne, is originated from pre-existing CNO nuclei, it depends on the initial metallicity of the star. Therefore, the weak \textit{s}-process is considered to be of secondary nature and that could explain why we do not see a steep rise toward older ages where stars are less metallic. 

The relations of abundance ratios with [Fe/H] are represented in Fig. \ref{all_feh_color_age} with a color scale to show the ages. This plot manifestly represents the mixed influence of metallicity and age on GCE. For the $\alpha$ elements (Mg, Si, Ti), at a given [Fe/H] the abundances increase as age increases, as also shown in Fig. \ref{alpha_feh_ages}. The metal poor thin disk ([Fe/H]\,$\lesssim$\,--0.3\,dex) presents well mixed ages meanwhile the metal rich counterpart only has stars younger than $\sim$\,5\,Gyr. The \textit{h$\alpha$mr} stars are clearly older compared to thin disk stars at the same metallicity. The abundances of oxygen have much larger errors and thus the separation by ages is not as clear as for the other $\alpha$ elements. On the other hand, the Ca abundances do not show so clear separation between the different populations but still we can distinguish a general trend of increasing age and [Ca/Fe] as [Fe/H] diminishes. 

A second group of elements is formed by Al, Cu and Zn which present a similar morphology. In this case the ages increase in a "diagonal" way, not only as the [X/Fe] ratio does but also with the drop of metallicity rather than at a fixed metallicity as happened for $\alpha$ elements. For the \textit{s}-process elements Sr, Y, Zr and Ba we can see how the thin disk presents a very clear stratification of ages, also in a 'diagonal' way, but with ages decreasing as both [X/Fe] and metallicity increases. However, the thick disk stars present a wider range of abundance ratios despite its narrow range of ages. The lower abundances of \textit{s}-process elements in \textit{h$\alpha$mr} with respect to thin disk stars at the same [Fe/H] could be caused by their older ages. The peak of abundances of \textit{s}-process elements around solar metallicity is formed by the youngest stars in the sample. However, the steep decrease of Ba abundances at supersolar metallicities does not depend on age. Ce and Nd present more mixed ages in the [X/Fe]-[Fe/H] plane, probably due to the larger errors of those abundance ratios. Finally, the abundances of Eu also show higher values as age increases and metallicity decreases, similar to O, but the ages are more mixed due to the larger uncertainties on the abundance derivation of this element.

\section{Stellar age estimation using chemical abundances}\label{sec:clocks}

As we have seen in previous sections stellar dating using chemical clocks is based on the different galactic chemical evolution of some species. The different contribution to the chemical evolution of the Galaxy of the SNeII, SNIa, and the low-mass AGB stars residuals opens the door to the stellar dating using certain surface chemical abundances \citep{nissen16}. The work by \citet{dasilva12} was the first exploring the relation with age of abundances ratios of Y or Sr over Mg, Al, or Zn. More recently, \citet{nissen15,nissen16} found that ratios of [Y/Mg], [Y/Al] or [Al/Mg] are precise age indicators in the case of solar twins stars. These are the so called chemical clocks and have been studied in other samples of solar twins \citep{spina16,tucci-maia16} and very recently, in a bigger sample of stars within the AMBRE project \citep{titarenko19}. Moreover, these chemical clocks working over solar twin stars where confirmed using stars dated by asteroseismology \citep{nissen17}. However, \citet{feltzing17} and \citet{delgado17_nice} find that, when stars of different metallicities and/or effective temperatures are included, these simple correlations are not valid anymore. This can be clearly seen in Fig. \ref{all_age_color_feh} where the slopes of [X/Fe] ratios with age are shown for three metallicity groups. For instance, [Ca/Fe] has a mostly flat behavior with age for stars with --0.6\,$<$\,[Fe/H]\,$<$\,--0.2\,dex (black line) and --0.2\,$<$\,[Fe/H]\,$<$\,0.2\,dex (blue line) but the metal rich stars ([Fe/H]\,$>$\,0.2\,dex, red line) present a negative trend. This is in contrast with the other $\alpha$ elements in most of the metallicity bins, which show an increase of abundance ratios with age. Recent studies report a new kind of supernovae subclass called calcium-rich gap transients \citep{perets10} which can be an important contributor to the enrichment of Ca in the Galaxy, thus it might be the reason of the different evolution of Ca with time as compared to other $\alpha$ elements. Another interesting element is Cu, for which the age-abundance slope increases with metallicity. The case of most \textit{s}-process elements is also remarkable, with a totally different slope for metal-rich and metal-poor subgroups, especially for Ce and Nd. The higher dispersion of heavy-\textit{s} elements at younger ages as compared to light-\textit{s} elements is certainly caused by the different metallicity of the stars. In other words, it seems that the nucleosynthesis channels producing heavy-\textit{s} elements at younger ages have a stronger dependence on metallicity than those producing light-\textit{s} elements. On the other hand, for several elements (O, Mg, Si, Ti, Zn, and Sr) the mix of stars with different metallicities can add dispersion to the abundance-age relation but without a significant change in slope. 

To quantify the variations of the abundance-age slopes (for thin disk stars only) at different metallicity ranges we present in Fig. \ref{heavy_slopes} the slopes of neutron-capture abundance ratios ([X/Fe]) with age, as a function of \textit{s}-process contribution as shown for solar twins in Fig. 6 of \citet{spina18}. In order to have enough stars with low errors in age within each range we have selected bins with a 0.4\,dex width in [Fe/H] covering the full metallicity range of thin disk stars. The bin containing the solar twins (--0.2\,$<$\,[Fe/H]\,$<$\,0.2\,dex) presents a similar result to that of \citet{spina18} except that Nd does not show a correlation with age for solar metallicity stars. These authors showed how the dependence with age, in other words the slope, is larger (and negative) as the \textit{s}-process contribution increases, with \textit{r}-process elements such as Eu presenting a negligible correlation with age. The authors conclude that the \textit{s}-process production in the thin disk (where their solar twins belong) has been more active than the production of \textit{r}-process elements. However, we note that this might not hold for all the metallicity ranges. At super-solar metallicities stars in the thin disk have produced less Ce and Nd in recent times and Y and Zr seem to be produced at a lower rate than for solar metallicity stars. On the other hand, Sr, presents a rather constant negative trend with age for all the metallicity bins. Nevertheless, we note that these results must be taken with caution because of the low significance of the slopes in such small subsamples.

Using the sample of reliable ages described in the previous sections, we have studied the most significant relations for estimating the stellar age. The main goals are: 1) find the dependences explaining the spread of [X/Fe] vs. age found when stars within a large range of metallicities and effective temperatures are observed, and 2) offer the best relations possible for a precise estimation of stellar ages. Among all the chemical species presented in DM17, there are two special cases: Eu (with large errors) and ZrI (only available for cool stars). There is a large number of stars with these abundances unknown so we have not used these two species to estimate stellar ages. In a similar way, we have not considered oxygen due to the large errors in the abundances and the difficulty of deriving it as compared to other $\alpha$ elements.

\begin{figure*}
\centering
\includegraphics[width=1\linewidth]{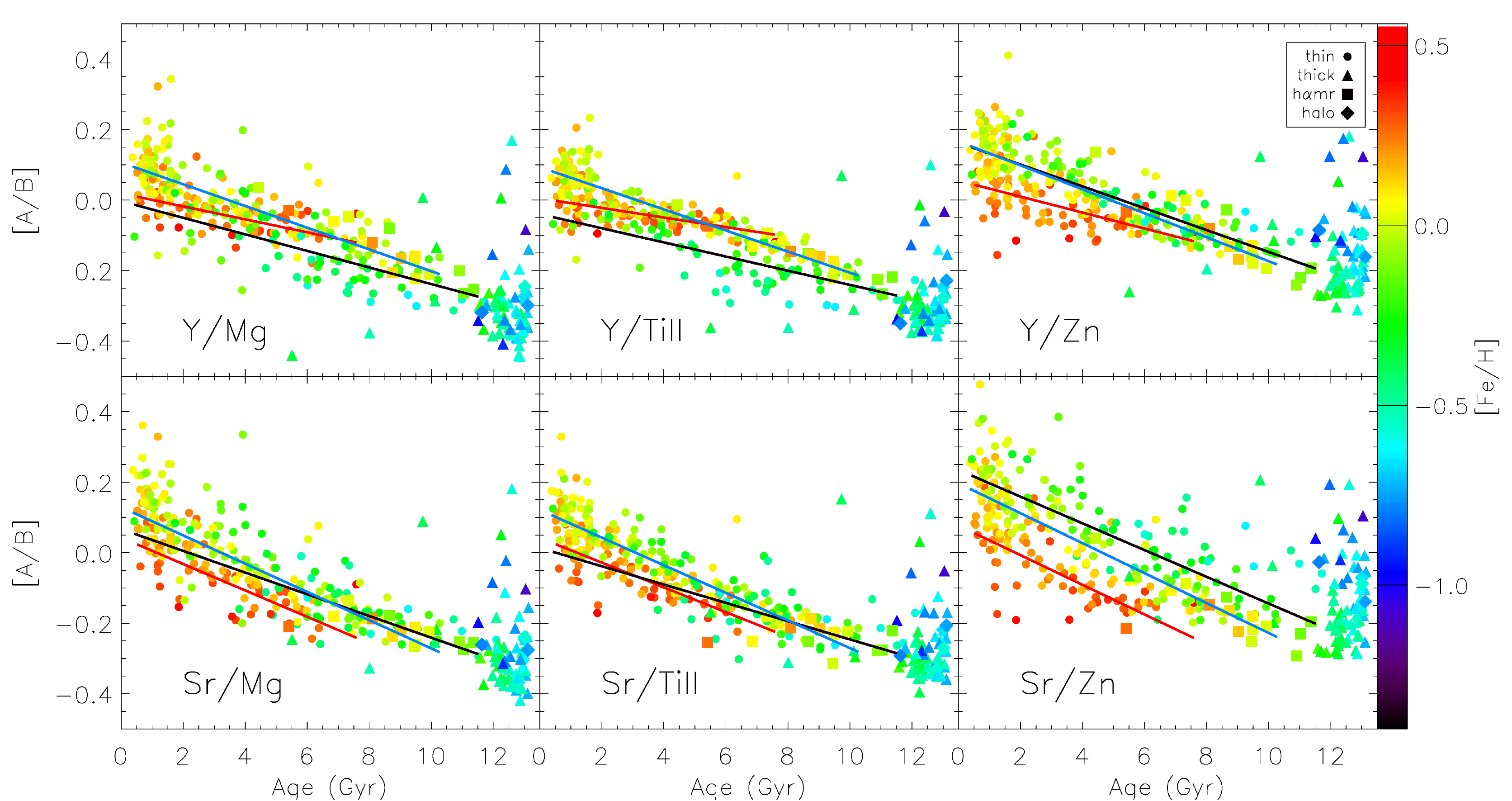}
\caption{Some of the chemical clocks studied as a function of age with the metallicities in color scale. The weighted linear fits are the same as in Fig. \ref{all_age_color_feh}.} 
\label{clocks_feh_color}
\end{figure*}

\begin{table}
\centering
\caption{Spearman correlation coefficients, $\rho$, of [X/Fe] abundance ratios vs. the stellar age, $T_{\rm eff}$, [Fe/H], and $M$ using the subsample with the most reliable stellar ages.}
\begin{tabular}{ccccc}
  \hline
Element & $T_{\rm eff}$ & [Fe/H] & $M$ & Age \\ 
  \hline
${\rm [O/Fe]}$    &   -0.29 &  -0.80 & -0.70 & 0.77\\ 
${\rm [Al/Fe]}$     &  -0.80 &-0.33 & -0.58 & 0.76\\ 
${\rm [Mg/Fe]}$     &  -0.55 &-0.70 & -0.73 & 0.89\\ 
${\rm [Si/Fe]}$     &  -0.38 &-0.72 & -0.59 & 0.81\\ 
${\rm [Ca/Fe]}$     &   -0.11 &-0.89 & -0.70 & 0.59\\ 
${\rm [TiI/Fe]}$    &  -0.32 &-0.81 & -0.67 & 0.72\\ 
${\rm [TiII/Fe]}$   &  -0.52 &-0.75 & -0.70  & 0.87\\ 
${\rm [Cu/Fe]}$    &   -0.43 &  0.45 & 0.18 & 0.13\\ 
${\rm [Zn/Fe]}$  &   -0.76 &-0.34  & -0.60 & 0.81\\ 
${\rm [Sr/Fe]}$    &    0.49 &-0.12 & 0.15 &  -0.43\\ 
${\rm [Y/Fe]}$      &   0.10 & 0.22 & 0.28 & -0.44\\ 
${\rm [ZrII/Fe]}$   &  0.03 &-0.56 & -0.36 &  0.15\\ 
${\rm [Ba/Fe]}$     &   0.29 &-0.16 & 0.04 & -0.33\\ 
${\rm [Ce/Fe]}$     &  0.01 &-0.16 & -0.04 & -0.14\\ 
${\rm [Nd/Fe]}$     &  -0.35 &-0.66 & -0.58 & 0.44\\ 
${\rm [Eu/Fe]}$   &  -0.39 &-0.76 & -0.72  & 0.70\\ 
   \hline
\end{tabular}
\label{tab:CC_age}
\end{table}

\begin{table}
\centering
\caption{Spearman cross-correlation coefficients, $\rho$, of potential chemical clocks vs. the stellar age, $T_{\rm eff}$, [Fe/H], and $M$ using the subsample with the most reliable stellar ages.}
\begin{tabular}{ccccc}
  \hline
Element & $T_{\rm eff}$ & [Fe/H] & $M$ & Age \\ 
  \hline
${\rm [Y/Mg]}$    &     0.47 & 0.62 & 0.68 &  -0.86 \\ 
${\rm [Y/Zn]}$    &     0.59 & 0.39 & 0.59 &  -0.80 \\ 
${\rm [Y/Al]}$    &     0.66 & 0.37 & 0.58 &  -0.78 \\ 
${\rm [Y/TiII]}$    &     0.45 & 0.66 & 0.66 & -0.86 \\ 
${\rm [Y/Si]}$    &     0.34 & 0.67 & 0.60 &  -0.79 \\ 
${\rm [Y/O]}$    &     0.33 & 0.74 & 0.70 &  -0.82 \\ 
${\rm [Sr/Mg]}$   &     0.65 & 0.50 & 0.66 &  -0.88 \\ 
${\rm [Sr/Zn]}$   &     0.74 & 0.18 & 0.48 &  -0.74 \\ 
${\rm [Sr/Al]}$   &     0.78 & 0.21 & 0.49 &  -0.73 \\ 
${\rm [Sr/TiII]}$   &     0.67 & 0.51 & 0.64 &  -0.88 \\ 
${\rm [Sr/Si]}$   &     0.64 & 0.42 & 0.57 & -0.83 \\ 
${\rm [Sr/O]}$   &     0.54 & 0.63 & 0.73 & -0.89 \\ 
${\rm [Y/(Mg + Si)]}$   &     0.42 & 0.62 & 0.65 & -0.84 \\ 
${\rm [Y/(Mg + Ti)]}$   &     0.45 & 0.64 & 0.67 & -0.86 \\ 
${\rm [Y/(Ti + Si)]}$   &     0.38 & 0.64 & 0.63 & -0.82 \\ 
${\rm [Y/(Mg + Ti + Si)]}$   &     0.42 & 0.64 & 0.65 & -0.85 \\ 
   \hline
\end{tabular}
\label{tab:CC_chem_clock}
\end{table}

\subsection{Revising the chemical clocks candidates}

The main characteristics of a chemical clock are its high correlation with age and, when a linear regression is fitted to it as a function of the stellar age, this regression is able to explain the main part of the observed variability. Most of the claimed chemical clocks in the literature fulfill these requirements. In this section we will analyze our best ages sampling trying to confirm and/or add new chemical clocks to those already known. To do so, we have studied the correlation between all the chemical species of our sample with age. In addition, and for reasons we describe in section \ref{sec:2d}, we have also analyzed the correlation of the chemical species with stellar mass, \teff\ and [Fe/H]. The correlation has been determined using the Spearman correlation coefficient ($\rho$). It measures the rank correlation between two variables and so depicts monotonic relationships.

\begin{figure*}
\centering
\setlength\fboxsep{0pt}
\setlength\fboxrule{0.25pt}
\fbox{\includegraphics[width=1\linewidth]{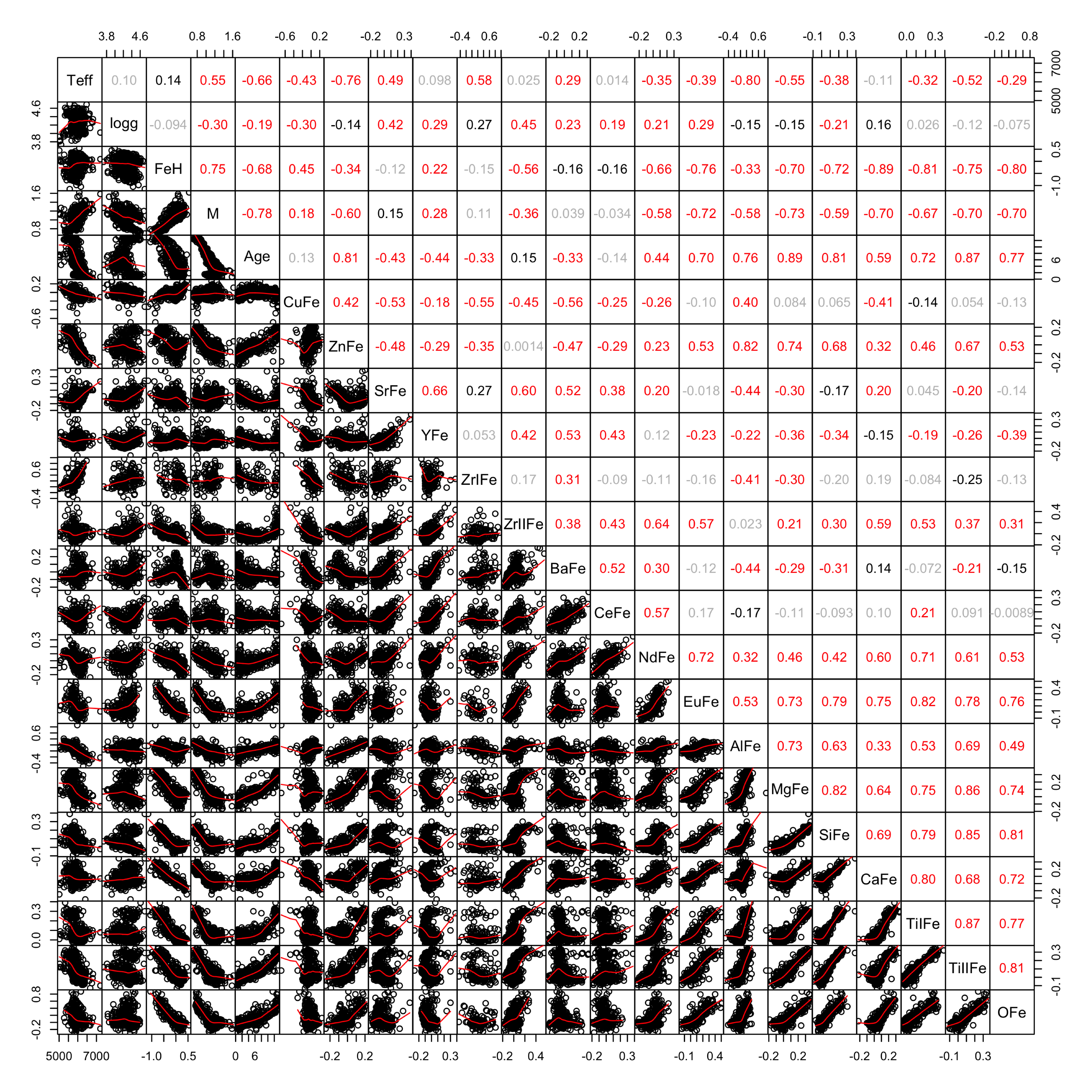}}
\caption{Correlations of all the variables with each other in the subset using the stars with the most accurate age determination (see text for details). In the upper triangular part, the Spearman statistic, $\rho$ for each correlation is depicted. The color of the font represents the statistical significance of the correlations depending on the p-value, with red, black, and gray for values with p\,$<$\,0.001, p\,$<$\,0.01 and p\,$<$\,0.1, respectively. In the lower triangular part we also show the spread-plot of each pair of variables, with LOESS fit just to guide the eye.}
\label{fig:corr}
\end{figure*}

In Table \ref{tab:CC_age} we show the values of $\rho$ found. We will first focus on the correlations with the stellar age. To ensure the statistical significance of our results, we have also obtained the p-value of each determination. In all the cases, except for those correlations lower than 0.1, the p-value is $<0.001$, that is, $\rho$ has a high statistical significance. We find that [Zn/Fe], [Al/Fe], [Si/Fe], [Mg/Fe], [TiI/Fe], [TiII/Fe], and [O/Fe] correlate with age ($|\rho| > 0.7$), [TiII/Fe] a little better than [TiI/Fe]. On the other hand, [Y/Fe] and [Sr/Fe] anticorrelate weakly with age, but they are the elements with the largest anticorrelation with age. Therefore, any ratio of [Y or Sr] over [Zn, Al, Si, Mg, or Ti] is a good candidate to be a chemical clock, that is, light-$s$ elements over $\alpha$ elements (plus Zn and Al). This is something we could expect from the chemical elements formation history. Some of these chemical clocks have already been studied. 

In Table \ref{tab:CC_chem_clock} we show the correlation of those ratios and the stellar age. Some of the clocks are shown as a function of age in Fig \ref{clocks_feh_color}. All chemical clocks candidates significantly anticorrelate with age with a p-value lower than 0.001. In other words, they are potentially good candidates for highly correlated 1D regressions with age, but some of them present a high dispersion caused by metallicity. We will verify this in the next section. With the information of Tables \ref{tab:CC_age} and \ref{tab:CC_chem_clock} we find some abundance ratios with a really high correlation with age, around $|\rho| > 0.8$. In the next sections we will study how to estimate stellar ages with abundance ratios. We also checked if we could get an improvement in the significance of the correlation with age\footnote{We also tried the ratios of Sr with combinations of $\alpha$ elements but the results were worse.} by using a combination of $\alpha$ elements with Y abundances (e.g., [Y/(Mg+Si)]). The $\rho$ values shown at the end of Table \ref{tab:CC_chem_clock} are similar to those for chemical clocks with a single $\alpha$ element and for simplicity we decided not to use them anymore.

For completeness, in Fig. \ref{fig:corr} we show all the correlations found using the best ages subset (upper triangle). The value of the correlation represents the Spearman correlation coefficient ($\rho$). The color of the font represents the statistical significance of the value of $\rho$ depending on the p-value, with red representing those with a p-value lower than 0.001, that is, the value has a very high statistical significance. All the correlations with $|\rho|>0.5$ have a p-value $<0.001$. In this figure we also show the spread-plot of each pair of variables, with a fitting using a LOESS (local polynomial regression) curve to guide the eye (bottom triangle). As expected, the elements produced by similar processes/stars are highly correlated. For example, all the $\alpha$ elements and Eu are correlated among them with $|\rho|\gtrsim0.7$. Zn is highly correlated with Al, Mg, Si, and TiII since all those elements have an important contribution from SNeII. However, the \textit{s}-process elements are not so correlated among them as the $\alpha$ elements, with $|\rho|$ values closer to $\sim$0.5. Finally, Nd is highly correlated with Eu, being both of them produced by the \textit{r}-process. However, only $\sim$50\% of Nd is produced in such process, which takes place in massive progenitors. The other half is produced by the \textit{s}-process in lower-mass stars. That would explain the fact that Nd is correlated with Ca, TiI and TiII (with $|\rho|>0.6$) since those $\alpha$ elements are both produced in SNeII and SNIa (i.e., higher and lower mass progenitors).

\begin{table*}
\centering
\caption{Best 1D relations. "Formula" is the independent variable used for estimating the stellar age and adj-$R^2$ is the statistic for measuring the goodness of the regression. We also detail the standard deviation (S.D.) of the age estimations using the regression compared with the "real" values for the test sample. Mean rel is the mean relative difference in $\%$ between estimations and "real" values. The rest are the linear regression coefficients and their uncertainties. See text for details.}
\begin{tabular}{crrrrrrr}
  \hline
Formula & adj-$R^2$ & S.D. & Mean rel & $a$ & $b$ & $\Delta a$ & $\Delta b$ \\ 
  \hline
        &           & Gyr &  & Gyr & Gyr/dex & Gyr & Gyr/dex \\ 
  \hline
  ${\rm [Y/TiII]}$ & 0.80 & 1.84 & 0.43 & 2.85 & -26.89 & 0.13 & 0.77 \\ 
  ${\rm [Y/Mg]}$ & 0.79 & 1.85 & 0.57 & 3.32 & -23.93 & 0.13 & 0.71 \\ 
  ${\rm [TiII/Fe]}$ & 0.79 & 1.89 & 0.53 & 3.37 & 33.87 & 0.13 & 1.01 \\ 
  ${\rm [Mg/Fe]}$ & 0.79 & 1.91 & 0.61 & 3.84 & 29.85 & 0.12 & 0.89 \\  
  ${\rm [Sr/Ti]}$ & 0.78 & 2.05 & 0.41 & 3.04 & -26.94 & 0.13 & 0.83 \\ 
  ${\rm [Sr/TiII]}$ & 0.77 & 2.13 & 0.63 & 3.34 & -24.58 & 0.13 & 0.78 \\ 
  ${\rm [Y/Ti]}$ & 0.76 & 1.61 & 0.57 & 2.68 & -27.75 & 0.15 & 0.90 \\ 
  ${\rm [Sr/Mg]}$ & 0.75 & 2.05 & 0.52 & 3.73 & -21.92 & 0.13 & 0.72 \\ 
  ${\rm [Ti/Fe]}$ & 0.74 & 1.94 & 0.67 & 3.16 & 35.50 & 0.14 & 1.21 \\ 
  ${\rm [Si/Fe]}$ & 0.68 & 2.28 & 0.56 & 3.17 & 45.66 & 0.16 & 1.79 \\ 
  ${\rm [Y/Zn]}$ & 0.68 & 2.40 & 0.62 & 5.12 & -26.55 & 0.13 & 1.06 \\ 
  ${\rm [Zn/Fe]}$ & 0.66 & 2.40 & 0.61 & 6.24 & 34.83 & 0.14 & 1.45 \\ 
  ${\rm [Y/Si]}$ & 0.66 & 2.04 & 0.65 & 2.92 & -30.25 & 0.18 & 1.27 \\ 
  ${\rm [Sr/Si]}$ & 0.65 & 2.36 & 0.60 & 3.34 & -28.48 & 0.17 & 1.19 \\ 
   \hline
\end{tabular}
\label{tab:1d}
\end{table*}

\begin{table*}
\centering
\caption{Comparison of 1D relations with those in the literature for the solar twins subset. Formula is the dependent variable estimated using the stellar age. adj-$R^2$ is the statistic to measure the goodness of the regression. The rest are the linear regression coefficients and their uncertainties. See text for details.}
\begin{tabular}{crrrrrc}
  \hline
\noalign{\smallskip} 
Formula & adj-$R^2$  & a$'$ & b$'$ & $\Delta a'$ & $\Delta b'$  & Source\\ 
\noalign{\smallskip} 
\hline
\noalign{\smallskip} 
\multirow{ 4}{*}{${\rm [Y/Mg]}$} & 0.84 &   0.209 & -0.041 & 0.023 &  0.003 & This work\\
& & 0.175 & -0.0404 & 0.011 & 0.0019 & \citet{nissen15}\\
& & 0.170 & -0.0371 & 0.009 & 0.0013 & \citet{nissen16}\\
& & 0.150 & -0.0347 & 0.007 & 0.0012 & \citet{nissen17}\\
& & 0.176 & -0.0410 & 0.011 & 0.0017 & \citet{spina16}\\
& & 0.186 & -0.0410 & 0.008 & 0.0010 & \citet{tucci-maia16}\\

\noalign{\smallskip} 
   \hline
\noalign{\smallskip} 
\multirow{ 5}{*}{${\rm [Y/Al]}$}& 0.84 &   0.210 & -0.042 & 0.024 &  0.004 & This work\\ 
&       & 0.196 & -0.0427 & 0.009 & 0.0014 & \citet{nissen16}\\
&       & 0.174 & -0.0400 & 0.008 & 0.0012 & \citet{nissen17}\\
&       & 0.194 & -0.0459 & 0.011 & 0.0018 & \citet{spina16}\\
\noalign{\smallskip} 
   \hline
\end{tabular}
\label{tab:1D_literature}
\end{table*}

\subsection{Multivariable linear regressions for the estimation of stellar ages}\label{sec:regressions}

One of the best options to condense all the information contained in a data sample in a simple relationship, with the aim of using it for stellar dating estimations in our case, is to obtain the most significant linear regressions possible. In fact, these relations try only to explain as much as possible the age variability. To avoid over-fitting and to keep the relations as simple as possible, we are going to present the study depending on the number of independent variables involved (or dimensions) up to the third dimension. We note that the linear regressions presented in next subsections are weighted by the errors on the variables (stellar parameters and abundance ratios).

\subsubsection{1D relations}

These relations estimate the stellar age using only one independent variable: Age $= f(X)$, where $f(X)=(a\pm \Delta a)+(b\pm \Delta b)\times X$ is the linear combination of the variable $X$. These are the relations we can usually find in the literature. In Table \ref{tab:1d} we show the relations with an adjusted $R^2$ (adj-$R^2)>0.65$, where adj-$R^2$ is a measurement of the dependent variable variance explained by the independent variable $X$ corrected by the number of dimensions involved. The p-value of all the adj-$R^2$ are, by far, lower than 0.001.

As expected, the relations with the largest adj-$R^2$ are those formed by the same abundance ratios with a large correlation with the age in Tables \ref{tab:CC_age} and \ref{tab:CC_chem_clock}. [Y/Mg], [Sr/Mg], and [Y/Zn] are part of the chemical clocks proposed by \citet{dasilva12,nissen15,nissen16,nissen17,spina16,delgado17_nice}. On the other hand, \citet{bensby14,haywood13,feltzing17} speculated that Ti can also be a good species for a chemical clock but with certain scatter. Here we show that, in fact, [Y/TiII] and [Sr/TiII] are very good ones. We also evaluated the chemical clocks with the average of TiI and TiII (named as Ti in the tables). Since the stars included in the sample are not only solar twins, the adj-$R^2$ obtained are not impressive because of the behavior of abundances ratios at different metallicities \citep{feltzing17,delgado17_nice}, but values of adj-R2 around 0.8 are remarkable.

To test the reliability of these regressions, we have randomly split the good ages subsample into a training and a control group, with a 70$\%$ and 30$\%$ of the stars respectively. We have re-obtained the regressions with the training group, estimated the ages for the testing group and compared them with the real ages (those in our catalog). In Table \ref{tab:1d} we show the standard deviation (S.D.) of the difference estimated - real ages. For these best relations, the S.D. is on the order of 2 Gyrs. On the other hand, since an error of 1 Gyr (for example) is not the same for a 13 Gyr old star or for a 1 Gyr old star in relative terms, we have also calculated the mean relative difference between the estimated ages and the real ages. This value (represented as a fraction of one), is shown in the column "Mean rel". In this case we can see that the mean error is, in fact, large for these 1D relations. The minimum error is on the order of 40$\%$. A number of additional consistency tests for the regression models shown in this work can be found at the Appendix \ref{sec:appendix_regressions}. In Fig. \ref{YTiII_clock_diffs} we also show a comparison example between the real ages and those obtained with the 1D formula using the [Y/TiII] chemical clock, only for a subset of stars with very low errors in age ($<$\,0.5\,Gyr). The results, shown with yellow circles, demonstrate a clear dependence of the obtained ages with the stellar parameters [Fe/H] and \teff, and the need to include more variables in the formulas to obtain the age (see next subsection).

We have also compared the relations substantiated by our database of solar twins with those found in the literature. To do so, we have selected a subset of stars with $T_{\rm eff}=[5677,5877] K$, and [Fe/H]$=[-0.1,0.1]$. We allow the errors in age to be up to 2\,Gyr to have enough stars for a meaningful comparison. In this case, the 1D relations are formulated as $X = f({\rm Age})$, so we have reproduced those in \citet{nissen15,nissen16,nissen17,spina16,tucci-maia16}. Our results compared with those in the literature are shown in Table \ref{tab:1D_literature}. All the coefficients in the literature and those we obtain are equivalent within uncertainties. We note, however, that we lack stars with low errors in age in the range 3-6\,Gyr and this might explain the small differences in the regression coefficients.

Finally, our data sample support no correlation for the expression [Al/Mg]$=f({\rm Age})$ as shown in \citet{nissen16}. This result is reasonable since both [Al/Fe] and [Mg/Fe] correlate positively with age, and, therefore, its ratio is not expected to be a good chemical clock, as its the case when we rate a chemical element correlating with another anticorrelating with age.

\begin{table*}
\centering
\caption{Best ten 2D relations with $T_{\rm eff}$ as independent variable. Formula are the independent variables  used for estimating the stellar age. "Rel imp X or $T_{\rm eff}$" is the relative importance of the variable $X$ or $T_{\rm eff}$ in the regression. The rest are the linear regression coefficients and their uncertainties. In columns "b" and "$\Delta$b", the form X(Y) represents $X\times 10^Y$.}
\begin{tabular}{lrrrrrrrrrrr}
  \hline
Formula & adj-$R^2$ & S.D. & Mean rel & Rel imp $T_{\rm eff}$ & Rel imp X & a & b & c & $\Delta a$ & $\Delta b$ & $\Delta c$ \\ 
  \hline
        &           & Gyr &  &  &  & Gyr & Gyr/K & Gyr/dex & Gyr & Gyr/K & Gyr/dex \\ 
  \hline
  $T_{\rm eff} +$ [Y/TiII] & 0.86 & 1.36 & 0.38 & 0.25 & 0.75 & 27.47 & -4.0(-3) & -23.33 & 2.33 & 4(-4) & 0.73 \\ 
  $T_{\rm eff} +$ [Y/Ti] & 0.85 & 1.61 & 0.33 & 0.27 & 0.73 & 33.15 & -5.0(-3) & -23.77 & 2.29 & 4(-4) & 0.76 \\ 
  $T_{\rm eff} +$ [Y/Mg] & 0.83 & 1.57 & 0.48 & 0.25 & 0.75 & 26.90 & -3.9(-3) & -20.70 & 2.54 & 4(-4) & 0.72 \\  
    $T_{\rm eff} +$ [Ti/Fe] & 0.81 & 1.87 & 0.67 & 0.26 & 0.74 & 25.97 & -3.8(-3) & 30.30 & 2.29 & 4(-4) & 1.07 \\ 
    $T_{\rm eff} +$ [Si/Fe] & 0.78 & 2.14 & 0.64 & 0.28 & 0.72 & 30.41 & -4.5(-3) & 37.50 & 2.35 & 4(-4) & 1.43 \\ 
  $T_{\rm eff} +$ [TiII/Fe] & 0.83 & 1.67 & 0.47 & 0.25 & 0.75 & 25.43 & -3.6(-3) & 29.57 & 2.63 & 4(-4) & 1.06 \\ 
  $T_{\rm eff} +$ [Y/Si] & 0.76 & 2.25 & 0.56 & 0.34 & 0.66 & 39.33 & -6.0(-3) & -24.06 & 2.35 & 4(-4) & 0.99 \\ 
  $T_{\rm eff} +$ [Sr/Ti] & 0.74 & 2.06 & 0.58 & 0.24 & 0.76 & 9.65 & -1.0(-3) & -23.39 & 3.10 & 5(-4) & 1.02 \\ 
 $T_{\rm eff} +$  [Sr/Mg] & 0.73 & 2.13 & 0.57 & 0.25 & 0.75 & 14.90 & -1.8(-3) & -19.83 & 3.06 & 5(-4) & 0.91 \\ 
  $T_{\rm eff} +$  [Y/Zn] & 0.66 & 2.25 & 0.72 & 0.31 & 0.69 & 26.93 & -3.4(-3) & -20.66 & 3.17 & 5(-4) & 1.18 \\ 

   \hline 
\end{tabular}
\label{tab:2d_teff}
\end{table*}

\begin{table*}
\centering
\caption{Best ten 2D relations with [Fe/H] as independent variable. Formula are the independent variables used for estimating the stellar age. adj-$R^2$ is the statistic to measure the goodness of the regression. "Rel imp X or [Fe/H]" is the relative importance (in fraction of unity) of the variable $X$ or [Fe/H] in the regression. The rest are the linear regression coefficients and their uncertainties.}
\begin{tabular}{lrrrrrrrrrrr}
  \hline
Formula & adj-$R^2$ & S.D. & Mean rel & Rel imp [Fe/H] & Rel imp X & a & b & c & $\Delta a$ & $\Delta b$ & $\Delta c$ \\ 
  \hline
        &           & Gyr &  &  &  & Gyr & Gyr/dex & Gyr/dex & Gyr & Gyr/dex & Gyr/dex \\ 
  \hline
  ${\rm [Fe/H]} +$ [Sr/Zn] & 0.87 & 1.44 & 0.32 & 0.51 & 0.49 & 4.65 & -7.98 & -16.07 & 0.09 & 0.28 & 0.57 \\ 
  ${\rm [Fe/H]} +$ [Sr/TiII] & 0.86 & 1.83 & 0.38 & 0.35 & 0.65 & 3.34 & -4.64 & -19.38 & 0.10 & 0.33 & 0.72 \\ 
  ${\rm [Fe/H]} +$ [Sr/Ti] & 0.86 & 1.59 & 0.41 & 0.35 & 0.65 & 3.13 & -4.47 & -21.23 & 0.11 & 0.34 & 0.79 \\ 
  ${\rm [Fe/H]} +$ [Sr/Mg] & 0.86 & 1.47 & 0.42 & 0.36 & 0.64 & 3.64 & -4.90 & -17.16 & 0.10 & 0.33 & 0.65 \\ 
  ${\rm [Fe/H]} +$ [Y/Zn] & 0.85 & 1.56 & 0.33 & 0.41 & 0.59 & 4.62 & -5.92 & -20.41 & 0.10 & 0.33 & 0.82 \\ 
  ${\rm [Fe/H]} +$ [Zn/Fe] & 0.84 & 1.53 & 0.48 & 0.42 & 0.58 & 5.46 & -6.18 & 27.03 & 0.10 & 0.33 & 1.09 \\ 
  ${\rm [Fe/H]} +$ [Sr/Al] & 0.83 & 1.57 & 0.46 & 0.52 & 0.48 & 4.35 & -7.97 & -12.30 & 0.10 & 0.32 & 0.53 \\ 
  ${\rm [Fe/H]} +$ [Y/TiII] & 0.81 & 1.33 & 0.47 & 0.32 & 0.68 & 2.98 & -1.43 & -24.34 & 0.13 & 0.50 & 1.14 \\ 
  ${\rm [Fe/H]} +$ [Y/Al] & 0.81 & 1.63 & 0.42 & 0.45 & 0.55 & 4.26 & -6.48 & -14.70 & 0.11 & 0.37 & 0.71 \\ 
  ${\rm [Fe/H]} +$ [Y/Mg] & 0.80 & 1.73 & 0.40 & 0.33 & 0.67 & 3.42 & -2.41 & -20.30 & 0.13 & 0.48 & 0.98 \\ 
   \hline
\end{tabular}
\label{tab:2d_metal}
\end{table*}

\begin{table*}
\centering
\caption{Best ten 2D relations with $M$ as independent variable. Formula are the independent variables used for estimating the stellar age. "Rel imp X or $M$" is the relative importance of the variable $X$ or $M$ in the regression. The rest are the linear regression coefficients and their uncertainties.}
\begin{tabular}{lrrrrrrrrrrr}
  \hline
Formula & adj-$R^2$ & S.D. & Mean rel & Rel imp $M$ & Rel imp X & a & b & c & $\Delta a$ & $\Delta b$ & $\Delta c$ \\ 
  \hline
        &           & Gyr &  &  &  & Gyr & Gyr & Gyr/dex & Gyr & Gyr & Gyr/dex \\ 
  \hline
  $M +$ [Sr/Ti] & 0.87 & 1.40 & 0.52 & 0.45 & 0.55 & 15.06 & -10.29 & -18.12 & 0.82 & 0.69 & 0.87 \\ 
  $M +$ [Sr/TiII] & 0.87 & 1.31 & 0.48 & 0.45 & 0.55 & 15.46 & -10.47 & -16.50 & 0.81 & 0.70 & 0.80 \\ 
  $M +$ [Y/TiII] & 0.87 & 1.47 & 0.30 & 0.43 & 0.57 & 13.52 & -9.04 & -18.75 & 0.89 & 0.75 & 0.91 \\ 
  $M +$ [Y/Mg] & 0.85 & 1.71 & 0.37 & 0.44 & 0.56 & 14.12 & -9.26 & -16.30 & 0.95 & 0.81 & 0.89 \\ 
  $M +$ [Sr/Mg] & 0.85 & 1.78 & 0.72 & 0.46 & 0.54 & 15.91 & -10.62 & -14.39 & 0.87 & 0.75 & 0.78 \\ 
  $M +$ [Y/Ti] & 0.85 & 1.54 & 0.54 & 0.45 & 0.55 & 14.63 & -10.06 & -18.34 & 0.94 & 0.79 & 1.02 \\ 
  $M +$ [Sr/Si] & 0.85 & 1.53 & 0.45 & 0.52 & 0.48 & 18.62 & -13.23 & -17.35 & 0.79 & 0.68 & 0.98 \\ 
  $M +$ [Y/Zn] & 0.84 & 1.38 & 0.56 & 0.50 & 0.50 & 18.90 & -12.51 & -16.24 & 0.81 & 0.73 & 0.97 \\ 
  $M +$ [Y/Si] & 0.83 & 1.83 & 0.62 & 0.52 & 0.48 & 17.95 & -12.82 & -18.11 & 0.88 & 0.75 & 1.14 \\ 
  $M +$ [Zn/Fe] &0.82 & 1.91 & 0.60 & 0.51 & 0.49 &19.4 & -12.4 & 20.7 & 0.80 & 0.80 & 1.30 \\
 
   \hline
\end{tabular}
\label{tab:2d_mass}
\end{table*}

\begin{table*}
\centering
\caption{Best ten 3D relations. Formula are the independent variables used for estimating the stellar age. "Rel imp $T_{\rm eff}$ or [Fe/H] or Z" is the relative importance of the variable $T_{\rm eff}$ or [Fe/H] or $X$ in the regression. The rest are the linear regression coefficients and their uncertainties. In column "b", the form X(Y) represents $X\times 10^Y$.}
\begin{tabular}{lllllllllllllll}
  \hline
Formula & $R^2$ & S.D. & Mean & RI & RI & RI & a & b & c & d & $\Delta a$ & $\Delta b$ & $\Delta c$ & $\Delta d$ \\
&&&rel&$T_{\rm eff}$&[Fe/H]&$X$&&&&&&&&\\
  \hline
        &           & Gyr &  &  &  &  & Gy & Gy/K & Gy/dex & Gy/dex & Gy & Gy/K & Gy/dex & Gy/dex \\ 
  \hline
  $T_{\rm eff} {\rm + [Fe/H]} +$ [Y/Zn] & 0.89 & 1.39 & 0.45 & 0.21 & 0.38 & 0.41 & 31.11 & -4.4(-3) & -6.62 & -13.93 & 2.38 & 4(-4) & 0.28 & 0.90 \\ 
  $T_{\rm eff} {\rm + [Fe/H]} +$ [Y/TiII] & 0.89 & 1.45 & 0.53 & 0.23 & 0.29 & 0.48 & 35.41 & -5.3(-3) & -3.71 & -15.88 & 2.25 & 4(-4) & 0.41 & 1.05 \\ 
  $T_{\rm eff} {\rm + [Fe/H]} +$ [Y/Ti] & 0.89 & 1.39 & 0.45 & 0.25 & 0.29 & 0.46 & 39.50 & -6.0(-3) & -3.85 & -15.90 & 2.13 & 3(-4) & 0.41 & 1.08 \\ 
  $T_{\rm eff} {\rm + [Fe/H]} +$ [Y/Mg] & 0.88 & 1.47 & 0.51 & 0.23 & 0.30 & 0.47 & 36.20 & -5.4(-3) & -4.41 & -13.05 & 2.28 & 4(-4) & 0.39 & 0.91 \\ 
  $T_{\rm eff} {\rm + [Fe/H]} +$ [Sr/Ti] & 0.88 & 1.59 & 0.40 & 0.20 & 0.33 & 0.47 & 23.84 & -3.4(-3) & -5.53 & -15.33 & 2.97 & 5(-4) & 0.35 & 1.12 \\ 
  $T_{\rm eff} {\rm + [Fe/H]} +$ [Sr/TiII] & 0.88 & 1.42 & 0.28 & 0.20 & 0.34 & 0.46 & 22.13 & -3.1(-3) & -5.60 & -14.34 & 3.14 & 4(-4) & 0.35 & 1.08 \\ 
  $T_{\rm eff} {\rm + [Fe/H]} +$ [Sr/Mg] & 0.88 & 1.09 & 0.40 & 0.20 & 0.35 & 0.45 & 24.16 & -3.4(-3) & -5.86 & -12.34 & 3.02 & 5(-4) & 0.34 & 0.93 \\ 
  $T_{\rm eff} {\rm + [Fe/H]} +$ [Y/Si] & 0.87 & 1.27 & 0.35 & 0.28 & 0.32 & 0.39 & 44.73 & -6.8(-3) & -5.25 & -14.61 & 2.11 & 3(-4) & 0.38 & 1.14 \\ 
  $T_{\rm eff} {\rm + [Fe/H]} +$ [Sr/Si] & 0.87 & 1.43 & 0.41 & 0.22 & 0.40 & 0.39 & 31.13 & -4.6(-3) & -6.83 & -13.55 & 2.87 & 5(-4) & 0.32 & 1.16 \\ 
  $T_{\rm eff} {\rm + [Fe/H]} +$ [Y/Al] &0.85 &1.67 &0.49 &0.22 &0.46 &0.32 &31.4 & -4.5(-3)& -7.4 & -7.7 &2.2 &4(-4) &0.3 &0.7 \\

  \hline
\end{tabular}
\label{tab:3d}
\end{table*}

\subsubsection{2D relations}\label{sec:2d}

The next natural step was to analyze the improvement we obtain when we add a second independent variable to the relation. Taking into account the lessons in the literature \citep[e.g.,][]{feltzing17} and in the previous sections of this work, one of the main sources for explaining the stellar age variability is the stellar metallicity. On the other hand, we cannot discard the impact of stellar structure and evolution on the transport of chemicals and hence on its chemical surface abundances \citep{salaris17,dotter17}. For example, diffusion is considered to have a major effect on hotter stars in the turn-off \citep[e.g.,][]{bertellimotta18} 
Therefore, we have tested all the possible combinations Age$=(a\pm \Delta a)+(b\pm \Delta b)*T_{\rm eff}+(c\pm \Delta c)*X$, with $X$ any of the chemical species or their ratios, and Age$=(a\pm \Delta a)+(b\pm \Delta b)*{\rm [Fe/H]}+(c\pm \Delta c)*X$. 

Stellar mass is, in terms of the physics involved in the stellar chemical mixing with age, a good proxy for stellar evolution. Thus, we have also used $M$ as an independent variable to obtain the age with a formula such as this: Age$=(a\pm \Delta a)+(b\pm \Delta b)*M+(c\pm \Delta c)*X$. When two variables have a large correlation, their inclusion in a linear regression is not recommended since they provide redundant information. In Tables \ref{tab:CC_age} and \ref{tab:CC_chem_clock} we show the correlations of all our chemical elements with \teff, [Fe/H], and $M$. We define that any pair of variables with a correlation with $|\rho|>0.7$ should not be used simultaneously in the 2D formulas. In Table \ref{tab:2d_teff} we show the final relations obtained for all the combinations of chemical elements and \teff\ with an adj-$R^2>$\,0.60. "Rel imp X or \teff" represent the relative importance of the independent variables in the explanation of the observed variance - that is, the real impact of each variable in the estimation of the stellar age. In this table we can see:
\begin{itemize}

\item The inclusion of \teff\ as additional variable increases adj-$R^2$ a $7.5\%$, going from 0.80 to 0.86 for the respective best cases. Therefore, with a chemical clock and the effective temperature as independent variables we can explain up to a $86\%$ of the stellar age variance.
\item The relative importance is almost always balanced, that is, both variables are needed for obtaining these results, with \teff\ explaining around a 25\% of the variance and the chemical clock the rest 75\%.
\item Seven of our proposed chemical clocks (all except [Sr/Al], [Sr/Zn], and [Y/Al]) have relations explaining at least a 64$\%$ of the stellar age variance.
\item Some [X/Fe] ratios can also be a good chemical clock when used together with the effective temperature. This is the case of [Si/Fe], [Ti/Fe], and [TiII/Fe].
\end{itemize}

Therefore, we can confirm that stellar structure and evolution plays a role in the observed stellar age variance when compared with surface abundances. We would like to note the decreasing in the S.D., compared with Table \ref{tab:1d}. The 2D relations provide a S.D. which is around 10-20 $\%$ more precise than 1D relations. Something similar happens in the case of the mean relative differences.

In Table \ref{tab:2d_metal} we show the best regression models obtained adding the stellar metallicity as second independent variable. In terms of the best statistics, the improvement reached compared with the 1D regressions is similar to that found in Table \ref{tab:2d_teff}. The chemical clocks involved in the best relations are also similar as using \teff\ as independent variables but the relative importance of [Fe/H] is higher than for \teff. However, in this case, the best relations (ordered by adj-$R^2$) are obtained with chemical clocks formed by Sr and Zn. We note that [Zn/Fe] and [Sr/Fe] already show a correlation with \teff\ (see Table \ref{tab:CC_age}), so combining them with the [Fe/H] improves the determination of ages. On the other hand, the abundances ratios [Ti/Fe] and [Si/Fe], which have a strong correlation with [Fe/H] (see Table \ref{tab:CC_age}), are better combined with \teff\ to obtain good 2D relations. Therefore, stellar metallicity also plays a role in the explanation of the stellar age variability when using chemical clocks.

Finally, using $M$ as independent variable also provides similar results. The increase in terms of adj-$R^2$ with respect to the 1D relations is on the order of previous 2D relations and the S.D. values are lower than when using \teff. The relative importances are even in a better balance than previously, probably because the stellar mass has a dependence on \teff\ and [Fe/H]. In addition, these results show a possible link between stellar structure and surface abundances for explaining the observed variability of stellar ages. This points in the direction of different theoretical studies \citep{dotter17,salaris17} since stellar mass is a rough proxy for how surface abundances change with age.

\subsubsection{3D relations}

As a natural last step, we analyzed the improvements reached when taking into account [Fe/H], \teff, and $M$ in groups of two, together with a stellar clock for estimating stellar ages. We have used relations with the form Age$=(a\pm \Delta a)+(b\pm \Delta b)*X+(c\pm \Delta c)*Y+(d\pm \Delta d)*Z$, with $X$ and $Y$ two of [Fe/H], \teff\, and $M$, and $Z$ any of the abundance ratios. We have followed the same procedure as in the previous cases, with the best ages subset. In this case, the results show that for every $Z$, any combination of [Fe/H], \teff\, and $M$ provides almost the same results in terms of adj-$R^2$. As an example, in Table \ref{tab:3d} we show the best results obtained in case of using [Fe/H] and \teff\ as independent variables. Here we can see that the adj-$R^2$ is on the order (just slightly larger) of that of the 2D relations. The same results are found when using [Fe/H] and $M$, and \teff\ and $M$. Our conclusion is that using a good proxy for describing stellar evolution, that counts for chemical mixing, contains almost all the information complementing that coming from the chemical clock. The remaining variability should come from another source and adding more general stellar characteristics does not add almost new information. 

Besides these considerations, other conclusions of this table are:

\begin{itemize}
\item The best 3D relations present a remarkable adj-$R^2$ = 0.89.
\item All the chemical clocks except [Sr/Al] and [Y/Al] (with an adj-$R^2$ close to our threshold) explain at least a 87$\%$ of the stellar age variance.
\item An important consistency check of these expressions is that the relative importance of the different variables are in general a 40$\%$ for the chemical clock, a 20$\%$ for the effective temperature, and a 40$\%$ for [Fe/H], in other words, well balanced.
\item The S.D. presents, in general, an improvement of a 30$\%$ with respect to the 1D relations. Something similar happens with the mean relative differences.
\end{itemize}

Therefore, as a general conclusion, a chemical clock in conjunction with one stellar parameter (either stellar mass or the effective temperature or the metallicity) or with the combination of both the effective temperature and the stellar metallicity can explain between a 89$\%$ and a 87$\%$ of the stellar age spread. This result confirms that stellar structure and evolution plays a role in stellar dating using surface chemical abundances, but we might need to consider additional physical processes, such as rotation, to understand the remaining variance. However, we note that most of our stars are slow rotators \citep[v \textit{sin}i\,$\lesssim$\,8 \kms,][]{delgado15}. Additionally, there are other factors not considered here, such as NLTE corrections or the atomic diffusion of elements with age that can affect the presented relations. Finally, we note that part of the spread observed in the [X/Fe] ratios is not astrophysical, and thus cannot be characterized. For example, the scatter in the abundance ratios has a dependence on the number of lines used for a given element \citep{adibekyan15c} and is, of course, affected by the errors in the stellar parameters. This is clearly shown by the low scatter of [X/Fe] values when using good quality spectra of solar twins with low errors in parameters \cite[e.g.,][]{jonay10}.

\subsubsection{Applicability}

\begin{figure}
\centering
\includegraphics[width=1\linewidth]{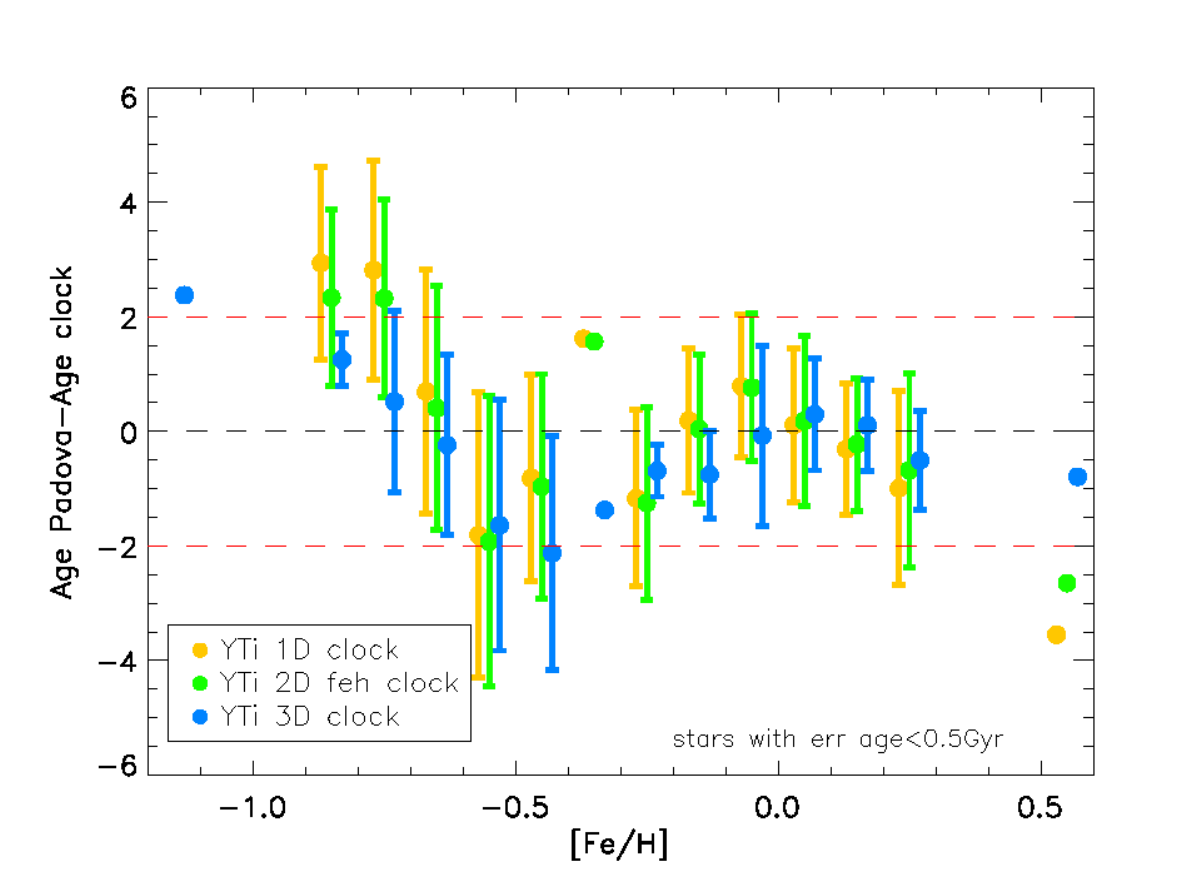}
\includegraphics[width=1\linewidth]{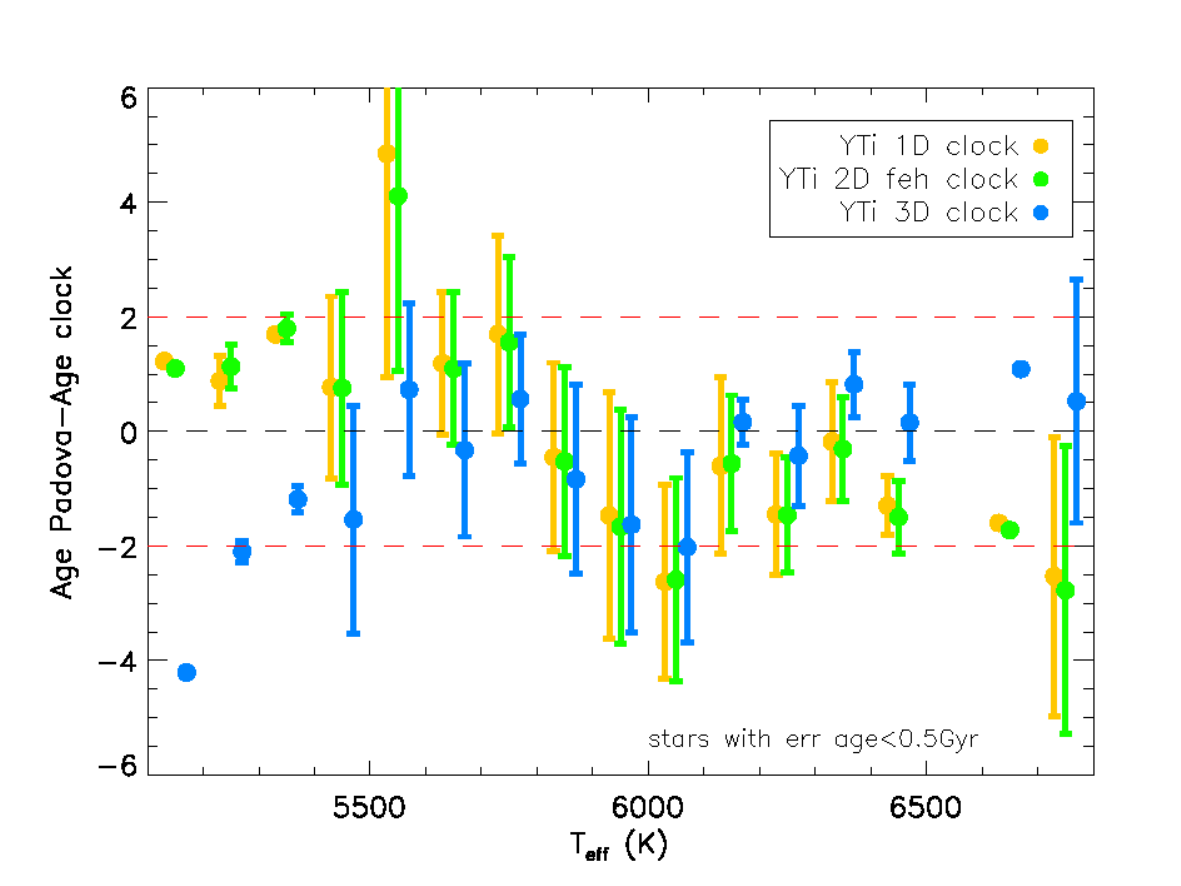}
\caption{Comparison of the [Y/TiII] 1D, 2D (using  [Fe/H]) and 3D clocks with the Padova ages for a small sample of stars with error in age lower than 0.5 Gyr. Average differences between age from formula and real age in different [Fe/H] bins (upper panel) and different \teff\ bins (lower panel). The dispersion of the differences are shown with the error bars when there are more than one star in a given bin.} 
\label{YTiII_clock_diffs}
\end{figure}

An ideal way to test if the formulas presented in previous subsections are reliable is to compare the ages we obtain from them with the real ages of the stars. However, our derived ages using isochrones can have large errors, especially in the case of cool stars. In Fig. \ref{YTiII_clock_diffs} we show how the average differences between ages obtained with the formulas (for the [Y/TiII] clock) and those derived by isochrones depend on \teff\ and [Fe/H]. In order to calculate the differences in ages we have used only those stars with errors in age lower than 0.5\,Gyr, that is, stars for which the isochrone ages are the closest possible to real within our sample. From the upper panel in Fig. \ref{YTiII_clock_diffs} it is clear that for metal-poor stars the differences can be quite large but for stars more metallic than --0.5\,dex the ages retrieved from the empirical relation are quite close to the real age. The ages obtained with the 3D formula (blue circles) are in better agreement with the derived ages in most of the metallicity bins. On the other hand, if we look at the applicability as a function of \teff\ (bottom panel), it is obvious that the 3D formula (which uses \teff) does not work well for cool stars. This is due to the fact that the linear regressions have been calculated using stars mostly hotter than 5300\,K. Therefore, we do not recommend using those formulas for cool stars, nor for stars hotter than 6500\,K due to the low number of hot stars used to derived the formulas. In a similar way, we do not recommend to use the formulas for metal poor stars since the regressions are obtained using stars mostly with [Fe/H]\,$>$\,--0.8\,dex. In most of the \teff\ bins the use of 3D clocks provides the most similar ages except for the stars around 6000\,K for which the formulas tend to overerestimate the age. 

\begin{figure}
\centering
\includegraphics[width=1\linewidth]{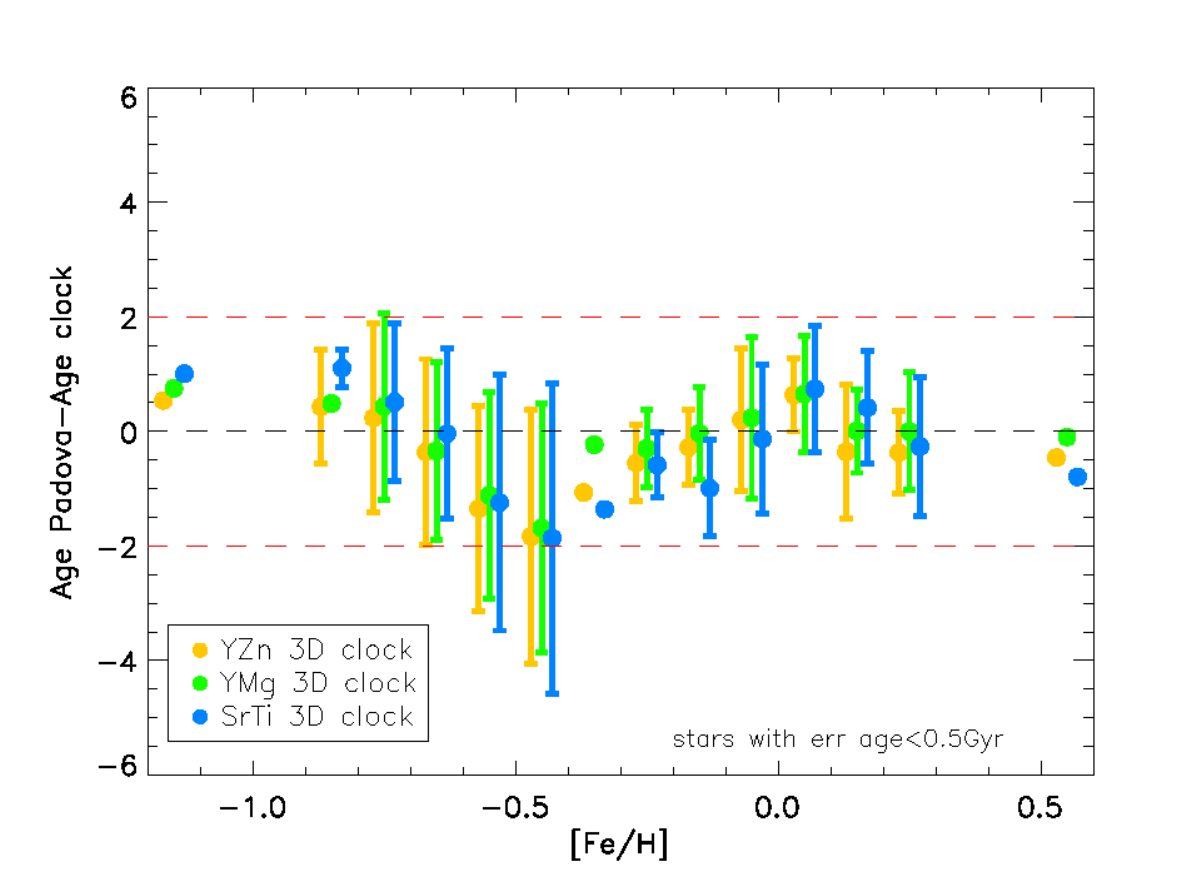}
\includegraphics[width=1\linewidth]{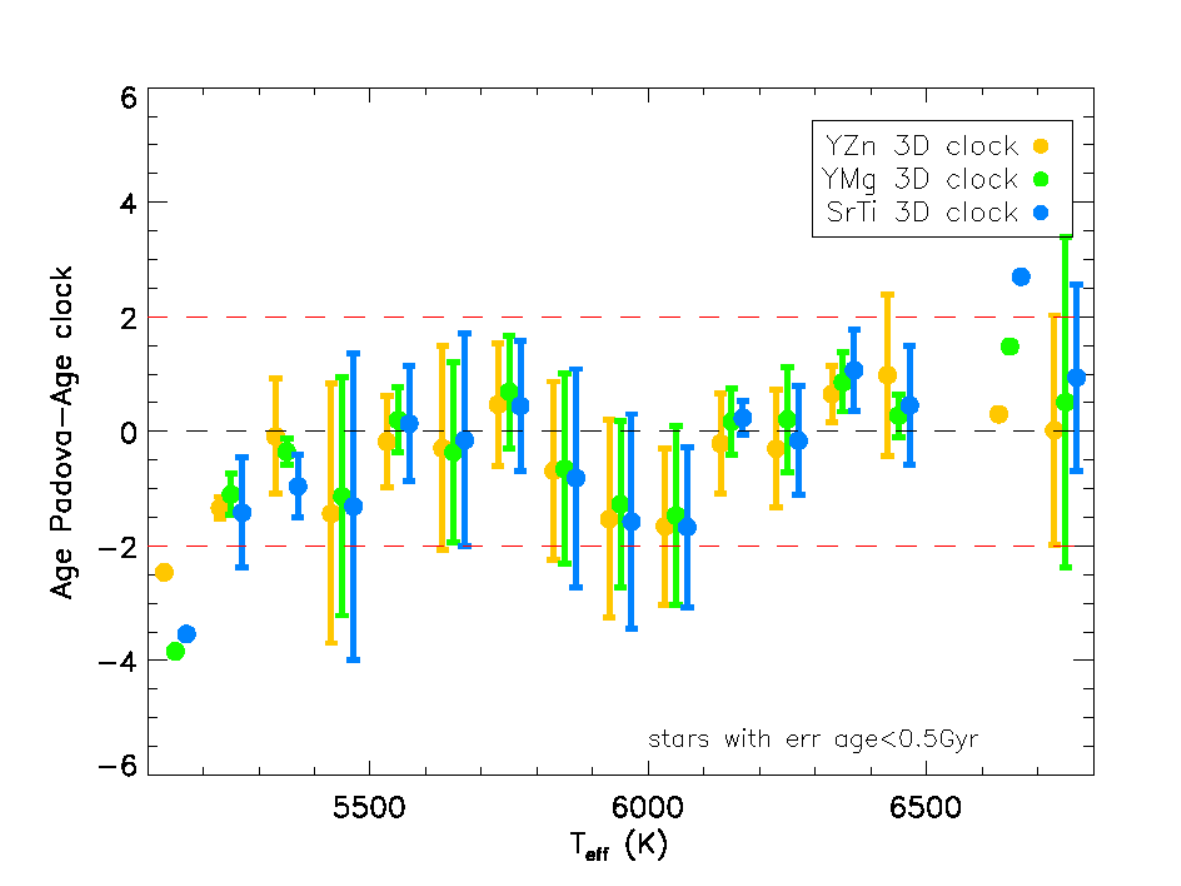}
\caption{Comparison of the 3D formulas using the chemical clocks [Y/Zn], [YMg], and [Sr/Ti] with the Padova ages for a small sample of stars with error in age lower than 0.5 Gyr. Average differences between age from formula and real age in different [Fe/H] bins (upper panel) and different \teff\ bins (lower panel). The dispersion of the differences are shown with the error bars when there are more than one star in a given bin.} 
\label{3Dclock_diff}
\end{figure}

In Fig. \ref{3Dclock_diff} we have compared the results obtained by using 3D formulas with different abundance ratios (and \teff\ and [Fe/H]). The differences and dispersions in a given bin are very similar among the different formulas, hence the different formulas of the same dimension provide comparable results. However, when deriving ages for a given star using formulas with different dimensions or parameters the results might not be similar. As a rule, we would recommend use of the 3D formulas or the 2D formula with the stellar mass, since they provide the best results. From the bottom panel of Fig. \ref{3Dclock_diff} we see that the ages from those three formulas are slightly better than the age from formula with [Y/TiII] for stars around 6000K (bottom panel of Fig. \ref{YTiII_clock_diffs}). Therefore, if several abundances ratios are available one can choose to get the ages from different formulas and get an average value. We also recall that when using these formulas for a given sample, the errors on the abundances of such sample (together with the errors in the coefficients) must be considered in order to have realistic errors on the determined age.

\section{Summary and conclusions}\label{sec:conclusions}

In this work we derived different set of ages using either parallaxes from \textit{Gaia} DR2 and Hipparcos or spectroscopic \logg\ for a sample of more than 1000 solar neighborhood stars belonging to the HARPS-GTO program. The chemical abundances of those stars were presented in previous works. The aim of this work is twofold. On the one hand we evaluated how abundance ratios of elements with different nucleosynthetic origin evolve with time. On the other hand we have provided different empirical relations to determine ages from abundance ratios and stellar parameters. The main results of this work can be summarized as follows. 

\begin{itemize}
 \item Our results confirm the large dispersion of the age-metallicity relation in the solar neighborhood, which increases with age except for stars older than 12\,Gyr, belonging to the thick disk. As found in previous works, the most metallic stars in our sample are not young, indicating that they had time to migrate to the solar neighborhood. The use of [$\alpha$/Fe] (being $\alpha$ the average of Mg, Si, and Ti) or [O/Fe] provides a much tighter relation with age which becomes steeper for thick disk stars. On the other hand, we found that [Zn/Fe] presents a single tight relation with age valid for both thin disk and thick stars. This is because thick disk stars are less enhanced in Zn than $\alpha$ elements with respect to thin disk stars.
 
 \item We find that the ages of thin disk stars in our sample reach up to 11 Gyr in few cases. The oldest thin disk stars have a similar age and [$\alpha$/Fe] content as the oldest \textit{h$\alpha$mr} but with a lower metallicity. These old thin disk stars have been regarded as coming from the outer disk \citep{haywood13}. Stars of a given [Fe/H] increase their ages as they increase their [$\alpha$/Fe], thus, \textit{h$\alpha$mr} are well differentiated from thin disk stars of similar metallicities. When looking at \textit{s}-process elements, the ages decrease with both [Fe/H] and [X/Fe] increasing. The maximum peak of [X/Fe] ratios for \textit{s}-process elements is formed by the youngest stars which are those having solar metallicity.
 
 \item Thick disk stars present a stronger enrichment in $\alpha$ elements when compared to thin disk stars of similar age. This is also true for the \textit{r}-process element Eu and for the light \textit{s}-process elements. However, heavy \textit{s}-process elements show a lower level of enrichment in the thick disk compared to light-\textit{s} elements. Intermediate mass AGB stars or rotating massive stars have been identified as possible responsibles for the light-\textit{s} elements overproduction in the thick disk.
 
 \item The abundances of $\alpha$ elements, Al, Zn, and Eu with respect to Fe generally increase with age whereas \textit{s}-process elements over Fe diminish with age. The elements with a contribution of both massive stars and lower mass stars (such as Cu and Nd) show a rather flat behavior with age. However, the trends present large dispersions for some of the abundance ratios, mostly caused by the wide range of metallicities in each age bin. In addition, the slopes of some of the [X/Fe]-age relations change with [Fe/H] indicating the strong influence of metallicity for some nucleosynthesis channels. We find that such slopes have a remarkable change for metal-rich stars in the cases of [Ca/Fe], [Cu/Fe], [Ce/Fe] and [Nd/Fe]. The slopes of [Y/Fe], [ZrII/Fe] and [Ba/Fe] vs age become rather flat for metal rich stars whereas some ratios such as [Mg/Fe], [Si/Fe], [Ti/Fe], [Zn/Fe] and [Sr/Fe] have a quite constant trend with age regardless of the metallicity.
 
 \item The observed variation of abundance-age slopes proves that the use of simple linear functions to derive ages from certain abundance ratios (also called chemical clocks) is limited to certain ranges of metallicity. Therefore, we investigate how the inclusion of stellar parameters in multivariable linear regressions can help in estimating stellar ages. Our results shows that by using different chemical clocks combined with one or two stellar parameters (\teff, [Fe/H] or stellar mass) we can explain up to a 89\% of the age variability. The derived formulas can thus be used as an age-proxy for stars for which the derivation of stellar ages through other methods is not possible. We note, however, that the empirical relations presented in this work have limitations and should not be used for stars outside the parameters range of our sample.
 
\end{itemize}

The overall results of this work show how important is to add the stellar age information when studying the GCE although we must be aware of the uncertainties involved in the derivation of such ages. This is especially a disadvantage for cool stars, because the large errors on stellar parameters will always prevent to get precise ages when using the isochrone method. Future asteroseismic observations will provide more reliable ages for different kind of stars. This will allow testing and revising of the results presented here and recalibration of the formulas when necessary, making them useful in a broader parameter range.

\begin{acknowledgements}
E.D.M., V.Zh.A., M. T., N.C.S., and S.G.S. acknowledge the support from Funda\c{c}\~ao para a Ci\^encia e a Tecnologia (FCT) through national funds
and from FEDER through COMPETE2020 by the following grants UID/FIS/04434/2013 \& POCI--01--0145-FEDER--007672, PTDC/FIS-AST/7073/2014 \& POCI--01--0145-FEDER--016880 and PTDC/FIS-AST/1526/2014 \& POCI--01--0145-FEDER--016886. E.D.M. acknowledges the support by the fellowship SFRH/BPD/76606/2011 funded by FCT (Portugal) and by the Investigador FCT contract IF/00849/2015/CP1273/CT0003 and in the form of an exploratory project with the same reference. V.Zh.A., N.C.S. and S.G.S. also acknowledge the support from FCT through Investigador FCT contracts IF/00650/2015/CP1273/CT0001, IF/00169/2012/CP0150/CT0002 and IF/00028/2014/CP1215/CT0002 funded by FCT (Portugal) and POPH/FSE (EC). A.M. acknowledges funding from the European Union's Horizon 2020 research and innovation program under the Marie Sklodowska-Curie grant agreement No 749962 (project THOT). W.J.C. acknowledges support from the UK Science and Technology Facilities Council (STFC). Funding for the Stellar Astrophysics Centre is provided by The Danish National Research Foundation (Grant agreement no.: DNRF106). J.I.G.H. acknowledges financial support from the Spanish Ministry project MINECO AYA2017-86389-P, and from the Spanish MINECO under the 2013 Ram\'on y Cajal program MINECO RYC-2013-14875. A.C.S.F. is supported by grant 234989/2014-9 from CNPq (Brazil). Finally, we thank the anonymous referee, whose detailed review helped to improve this paper.\\

This work has made use of data from the European Space Agency (ESA) mission \textit{Gaia} (https://www.cosmos.esa.int/gaia), processed by the \textit{Gaia} Data Processing and Analysis Consortium (DPAC,https://www.cosmos.esa.int/web/gaia/dpac/consortium). Funding for the DPAC has been provided by national institutions, in particular the institutions participating in the {\it Gaia} Multilateral Agreement.
This research has made use of the SIMBAD database operated at CDS, Strasbourg (France).\\

The core analysis was performed with R version 3.3.1 \citep{R}, RStudio Version 1.0.143, and the R libraries dplyr 0.5.0 \citep{dplyr2016}, foreach 1.4.3 \citep{foreach2015}, corrplot 0.86 \citep{corrplot}, parallel \citep{R}, RcppParallel \citep{RcppParallel}, and relaimpo 2.2-2 \citep{relaimpo2015}.
\end{acknowledgements}

\bibliographystyle{aa}
\bibliography{edm_bibliography}


\appendix

\section{Consistency of the regressions}\label{sec:appendix_regressions}

We have developed a large number of tests to ensure the consistency and reliability of the regressions shown at Section \ref{sec:regressions}. We have selected randomly 36 of the relations shown in Tables \ref{tab:1d} to \ref{tab:3d} for illustrating these tests.

\begin{figure*}
\centering
\setlength\fboxsep{0pt}
\setlength\fboxrule{0.25pt}
\fbox{\includegraphics[width=1\linewidth]{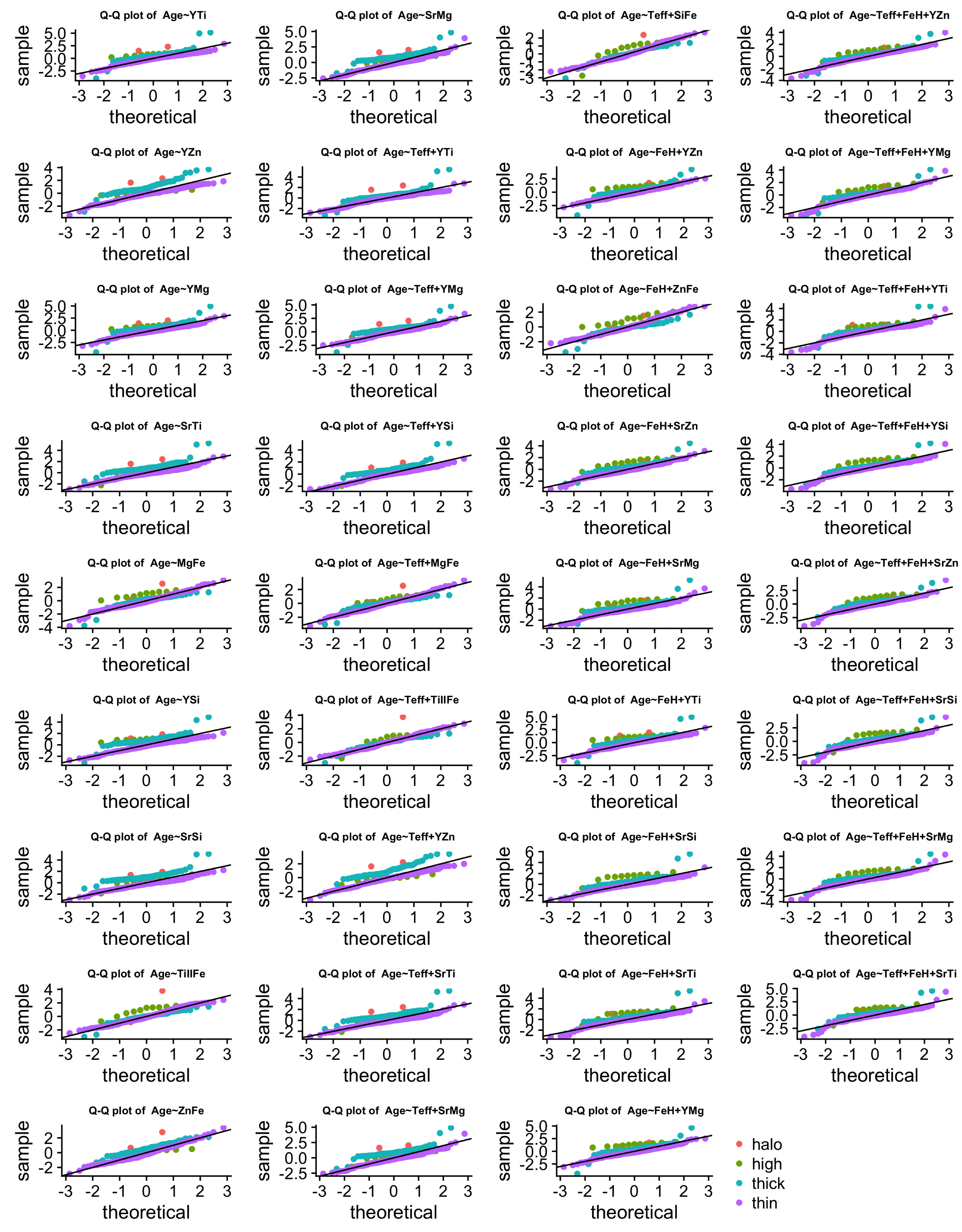}}
 \caption{quantile-quantile plots of the relations. See text for details.}
 \label{fig:Q-Q}
\end{figure*}

In Fig. \ref{fig:Q-Q} we show quantile-quantile plots of all the selected relations in a form to show whether the standardized residuals are normally distributed. The ordered standardized residuals are plotted on the ordinate of each plot, while the expected order statistics from a standard normal distribution are on the abscissa. Points close to the straight line are consistent with a normal distribution. The different classification of the stars as being part of the thin disk, thick disk, halo, or h$\alpha$mr is shown in different colors. In this figure we can see that all the relations, in general, follow this straight line. Therefore, the use of linear regression is justified, but we must explore the special case of relations taking into account only thin disk stars.

\begin{figure*}
\centering
\setlength\fboxsep{0pt}
\setlength\fboxrule{0.25pt}
\fbox{\includegraphics[width=1\linewidth]{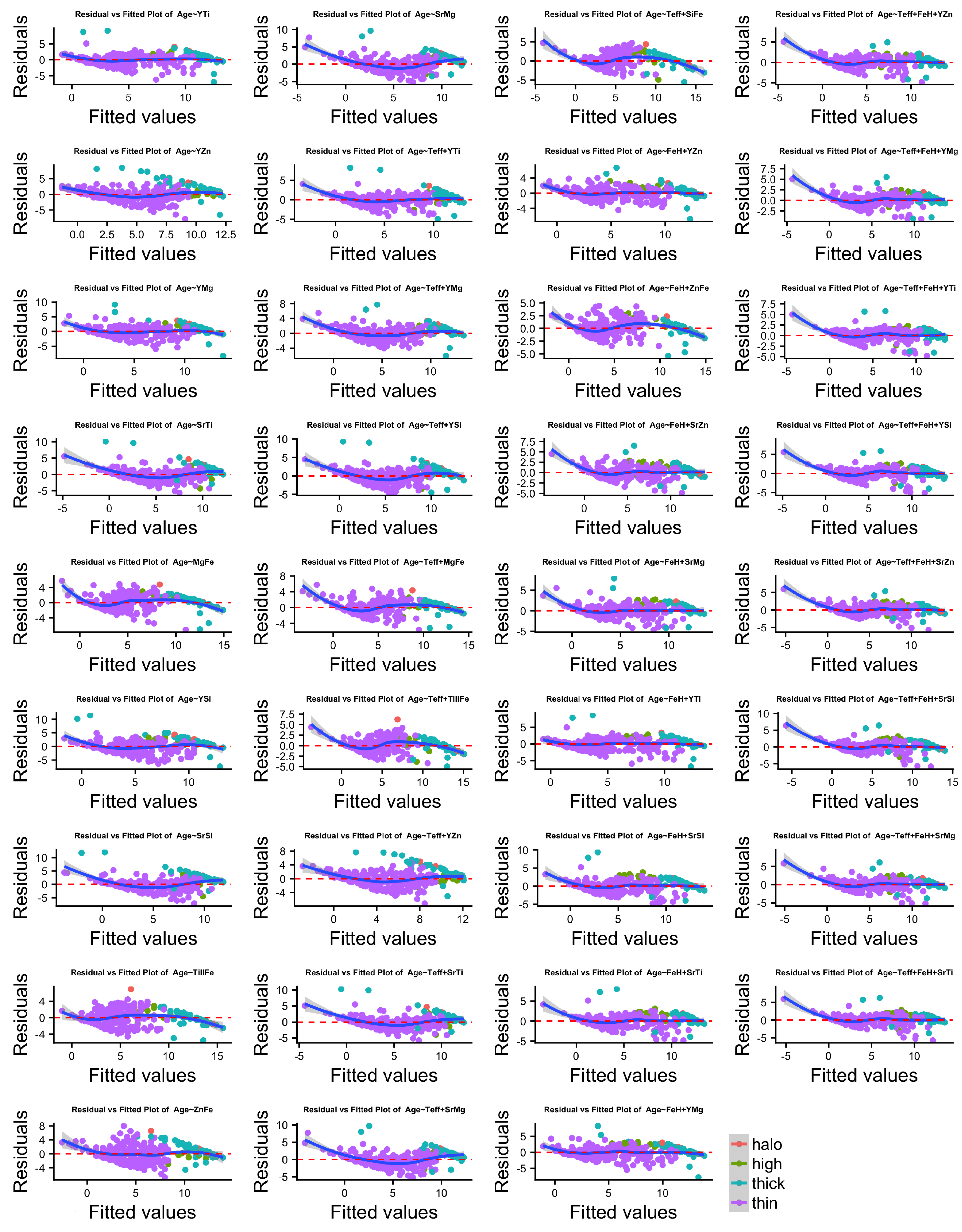}}
 \caption{Residuals vs. fitted values of the relations. The line is a LOESS curve to guide the eye. See text for details.}
 \label{fig:Residuals}
\end{figure*}

In Fig. \ref{fig:Residuals} we present the residuals as a function of the fitted values of these relations. Any clear trend in these residuals can be a signature of inaccurate or inefficient regression. We have also added a LOESS (local polynomial regression) curve to guide the eye. In this figure we can see that, in general, there are no clear trends in the distributions of the residuals. In every plot, the main body of points is randomly distributed around the value zero. We can only see boundary effects, where for the extreme age cases (close to zero and close to the limit of 13 Gyrs) the departure from the regressed model are larger. In addition, we can see in some cases a number of boundary stars with a large residual. 

\begin{figure*}
\centering
\setlength\fboxsep{0pt}
\setlength\fboxrule{0.25pt}
\fbox{\includegraphics[width=1\linewidth]{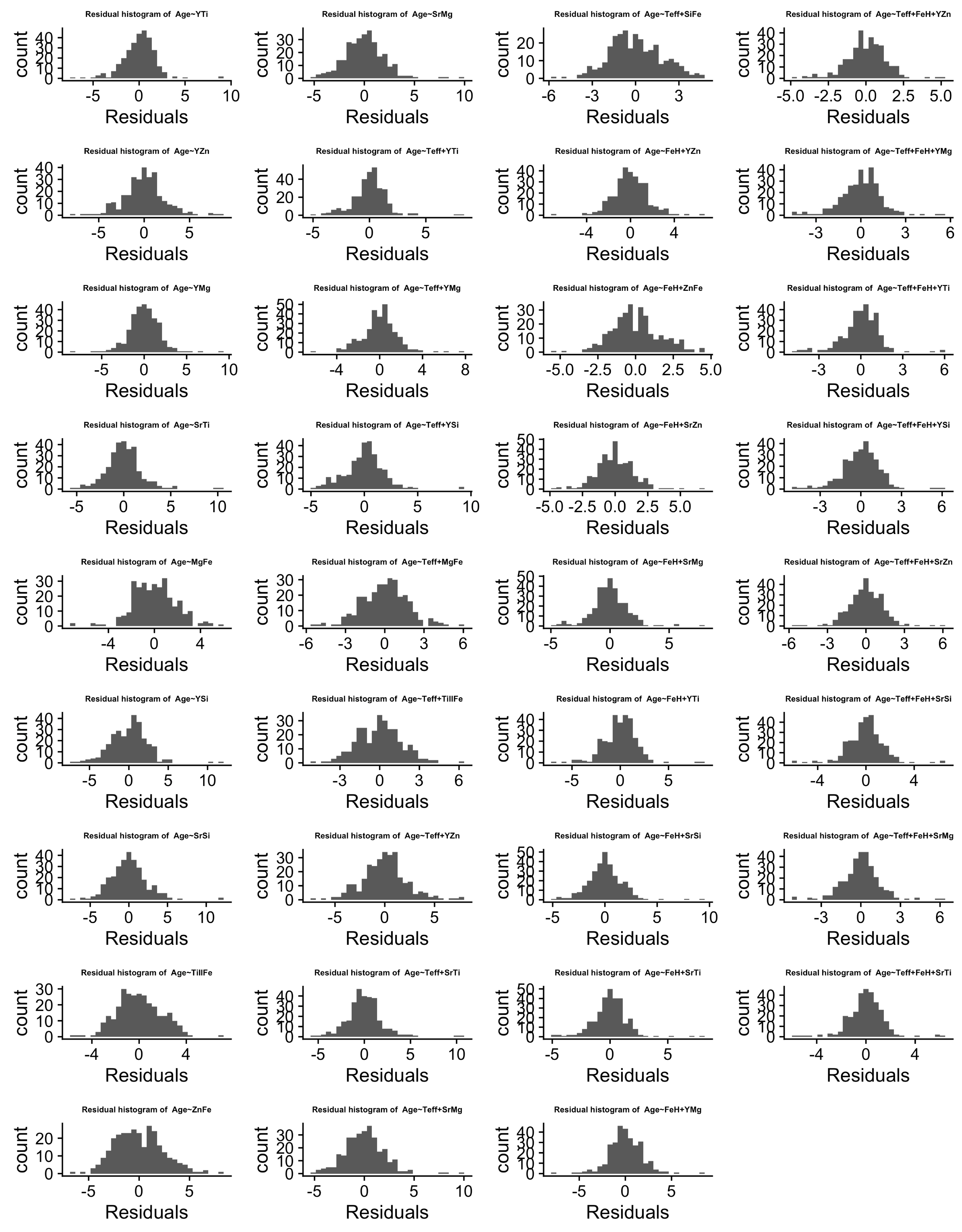}}
 \caption{Histogram of residuals of the relations. See text for details.}
 \label{fig:Residuals_hist}
\end{figure*}

Attending to the histogram of these residuals (Fig. \ref{fig:Residuals_hist}), we find that the distributions are mostly close to Gaussian-like distributions, fulfilling one of the main assumptions of regressions models.

\end{document}